\documentclass[twocolumn]{aastex701}

\newcommand\edit[2]{\added{#2}}
\newcommand\editcap[1]{#1}

\usepackage{amsmath}
\begin{document}

\title{A New Robust Constraint on the Self-interaction Cross-section of Dark Matter with Double Radio Relic Clusters}

\correspondingauthor{M. James Jee}

\author[orcid=0000-0002-5751-3697,gname=Myungkook James, sname=Jee]{M. James Jee}
\affiliation{Department of Astronomy, Yonsei University, 50 Yonsei-ro, Seoul 03722, Korea}
\affiliation{Department of Physics and Astronomy, University of California Davis, One Shields Avenue, Davis, CA 95616, USA}
\email[show]{mkjee@yonsei.ac.kr}

\author[gname=Dongak, sname=Park]{Dongak Park} 
\affiliation{Department of Astronomy, Yonsei University, 50 Yonsei-ro, Seoul 03722, Korea}
\email{dongaks@yonsei.ac.kr}

\author[orcid=0000-0002-1566-5094, gname=Wonki, sname=Lee]{Wonki Lee}
\affiliation{Department of Astronomy, Yonsei University, 50 Yonsei-ro, Seoul 03722, Korea}
\email{wonki.lee@yonsei.ac.kr}

\begin{abstract}
Merging galaxy clusters are a promising laboratory for measuring the self-interaction cross-section (SICS) of dark matter. However, previous studies have focused on galaxy-mass offsets, which numerical simulations have shown to be intrinsically small because galaxies remain tightly coupled to the dominant dark matter potential even with significant self-interaction. Their interpretation is further complicated by unknowns of the merger phase, geometry, and initial conditions.
In this paper, we overcome these obstacles by introducing the shock-to-shock distance, traced by double radio relics, as a merger chronometer that time-stamps the post-pericenter dynamical phase. Because the propagation speed of merger shocks is nearly independent of the SICS, while the halo-to-halo distance is depressed by SIDM-induced drag, the ratio of the two distances translates directly into a constraint on $\sigma/m$.
Applying this method to a gold sample of eleven cluster mergers hosting symmetric double radio relics, we determine \edit1{an upper limit on the SICS of $\sigma/m < 0.22~(0.63)~\text{cm}^{2}\,\text{g}^{-1}$ at the 68\% (95\%) confidence level}. This is the first constraint from cluster collisions that fully marginalizes over mass uncertainty, viewing angle, collision speed, merger phase, impact parameter, and gas profile slope.
\end{abstract}

\section{Introduction} 

The standard Cold Dark Matter (CDM) paradigm has been remarkably successful in explaining the large-scale structure of the universe. However, persistent discrepancies on small scales, such as the core-cusp and too-big-to-fail problems, suggest that dark matter (DM) may possess a non-zero self-interaction cross-section (SICS) \citep{Spergel2000, Tulin2018}. While these astrophysical puzzles provide a bottom-up motivation for self-interacting dark matter (SIDM), there is a strong top-down rationale from theoretical particle physics. Many dark matter candidates arise from extensions of the Standard Model in which they reside in a hidden sector with its own gauge symmetries; in such constructions, self-interactions mediated by hidden-sector gauge bosons are generic rather than exceptional \citep{Tulin2013}. Such models naturally predict self-interactions that can be many orders of magnitude larger than those between dark and visible matter, making the SICS a unique window into the physics of the dark sector \citep{Tulin2018}.

While terrestrial experiments continue to search for DM particles, colliding galaxy clusters serve as powerful astrophysical laboratories to probe these hidden-sector properties beyond the reach of terrestrial experiments. These merging clusters allow us to observe DM behavior during high-velocity impacts, where the separation of distinct cluster components provides a direct handle on DM physics \citep[e.g.,][]{Markevitch2004}. 

In a typical cluster merger, the hot intra-cluster medium (ICM) experiences ram-pressure stripping and lags behind the collisionless galaxies and DM halos. If DM is truly collisionless, the DM halos should coincide with the galaxies. Conversely, a significant SICS would induce a momentum-transfer drag, resulting in an offset between the DM mass centroid and the galaxies \citep{Markevitch2004, Harvey2015}. 

However, recent studies have demonstrated that these mass-galaxy offsets are susceptible to large systematic uncertainties that undermine their reliability as a dark matter probe \citep{Wittman2018}. The primary challenge arises from the complex dynamical behavior of galaxies that are not perfect tracers of the dark matter potential. Instead, galaxies often undergo significant sloshing motions within the dark matter halo on time scales much shorter than the overall cluster orbital period \citep{Kim2017}. This internal oscillation can produce transient offsets driven by local gravitational dynamics rather than by dark matter self-interaction. 

Furthermore, numerical simulations have shown that even in the presence of a non-zero cross-section, the expected mass-galaxy offset is frequently too small to be detected with current observational precision \citep{Kahlhoefer2014}. Because galaxies remain gravitationally bound to the dark matter halos, which provide the dominant contribution to the local potential, they tend to stay centered within the halo despite the momentum-transfer drag acting on the dark matter particles. This strong gravitational coupling means that a lack of observed offset does not necessarily imply a lack of self-interaction, leading to significantly weaker and more ambiguous constraints than previously assumed. 

The interpretation of these offsets is further plagued by the inherent ambiguity of the merger phase and the projection effects of the viewing angle \citep{Wittman2018}. A prominent example is the study by \citet{Harvey2015}, which initially claimed a stringent 95\% confidence-level upper limit of $\sigma/m < 0.47~\text{cm}^{2}\,\text{g}^{-1}$ by combining dark matter-galaxy offsets across 72 cluster collisions. However, subsequent reanalysis by \citet{Wittman2018} identified substantial methodological flaws, noting that averaging such offsets is physically unsound because their magnitude and direction depend critically on whether the system is in an approaching or receding phase. When these errors were rectified and the uncertainties properly marginalized, the constraint was relaxed by a factor of four. 

\begin{figure}
\includegraphics[width=8.5cm]{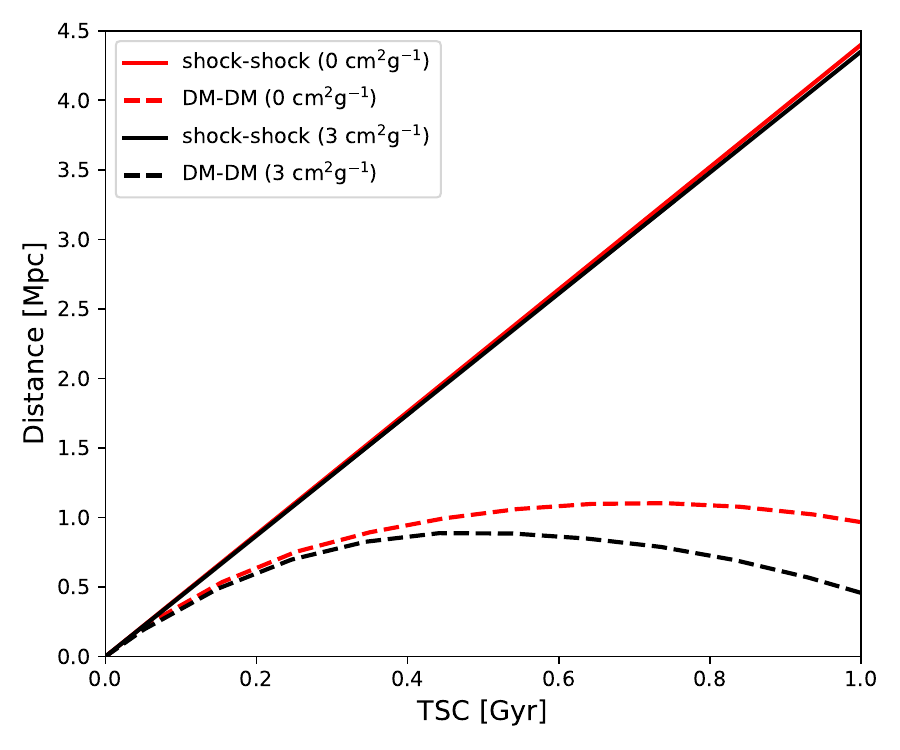}

\caption{Time evolution of the shock-to-shock and halo-to-halo distances. We simulate a merger of two $5\times10^{14}\,M_{\sun}$ halos and trace the halo-to-halo (dashed) and shock-to-shock (solid) distances as a function of the time since collision (TSC), with TSC$=0$ at pericenter passage. The shock-to-shock distance is nearly identical in the collisionless ($\sigma/m=0~\text{cm}^{2}\,\text{g}^{-1}$, red) and self-interacting ($\sigma/m=3~\text{cm}^{2}\,\text{g}^{-1}$, black) DM cases. However, the SIDM halos lose orbital energy via momentum-transfer drag, reaching apocenter earlier and at smaller separations. The halo-to-halo distance relative to the shock-to-shock distance is therefore a robust probe of self-interaction strength.
}
\label{fig:t_vs_d}
\end{figure}

In this paper, we introduce a new, powerful approach by utilizing the shock-to-shock distance relative to the halo-to-halo distance as the indicator of dark matter self-interaction.
While DM halos slow down following a collision due to self-interaction, the propagation speed of merger shocks remains nearly independent of the SICS (Figure~\ref{fig:t_vs_d}).
By comparing the shock-to-shock distance to the halo-to-halo distance derived from weak-lensing (WL) analysis, we can significantly reduce the geometric and temporal uncertainties that have plagued SICS measurements.

Double radio relics are diffuse radio sources that trace merger shock fronts, providing a direct measurement of the shock-to-shock distance \citep{Ensslin1998, Feretti2012, vanWeeren2011, vanWeeren2019}. They arise from diffusive shock acceleration (DSA) of relativistic electrons at outward-propagating merger shocks and appear as Mpc-scale, arc-like structures at the cluster outskirts. Systems hosting two relics on opposite sides of the cluster are especially useful: the symmetric geometry indicates a nearly head-on binary merger viewed close to the plane of the sky \citep{vanWeeren2011, Golovich2019}, and the relic pair directly marks the positions of both shock fronts, enabling a clean measurement of the shock-to-shock separation.

We apply this method to a sample of eleven ``gold" cluster mergers, which are selected for their double radio relics, clear mass bimodality, and near head-on collision geometry to minimize systematic modeling errors, providing the most stringent and marginalized constraint on the DM self-interaction cross-section to date from cluster-scale collisions: \edit1{$\sigma/m < 0.22~(0.63)~\text{cm}^{2}\,\text{g}^{-1}$ at the 68\% (95\%) confidence level}.
The paper is organized as follows. In \S\ref{sec:methods}, we describe our numerical simulations, Bayesian inference framework, and methodological validation with mock data. \S\ref{sec:results} presents the application to our gold sample and the resulting SICS constraint. Systematic effects and future prospects are discussed in \S\ref{sec:discussions}, and we summarize our findings in \S\ref{sec:conclusion}.

\section{Numerical Simulations and Methodology}\label{sec:methods}
\subsection{Cluster Collision Simulation Setup}\label{sec:simsetup}
To model the observable signals from merging clusters, we perform a suite of numerical hydrodynamic simulations using the GIZMO code (version 2022; \citealt{Hopkins2015}), which incorporates dark matter self-interaction between pairs of phase-space patches \citep{Rocha2013}. We generate initial conditions for merging halos using the {\tt cluster\_generator} package\footnote{\url{https://github.com/jzuhone/cluster_generator}}. The dark matter halo is assumed to follow a Navarro-Frenk-White (NFW) density profile \citep{NFW1997}:
\begin{equation}
    \rho_{dm}=\frac{\rho_{s,dm}}{(r/r_{s})(1+r/r_{s})^{2}}
\end{equation}
where $\rho_{s,dm}$ represents the scale density and $r_s$ is the scale radius defined by the virial radius $r_{200}$ and the concentration parameter $c$. The initial value of $r_{s}$ is determined through the mass-concentration relation of \citet{Duffy2008}. For the gas component, we utilize a beta profile:
\begin{equation}
    \rho_{gas}=\frac{\rho_{s,gas}}{[1+(r/r_c)^{2}]^{3\beta/2}}
\end{equation}
where the gas core radius $r_c$ is fixed at $r_c = r_s/2$. Scale densities are selected to achieve a fiducial gas fraction of 0.1 at $r_{500}$.

The collision kinematics are initialized by defining the relative velocity $v_{ini}$ between the two halos at their initial separation $d$, which is defined as the sum of their virial radii ($d = r_{200,1} + r_{200,2}$). This initial velocity is characterized by a velocity factor $f_{v}=v_{ini}/v_{ff}$, where $v_{ff}$ is the free-fall velocity evaluated at distance $d$ \citep{Sarazin2002}:
\begin{multline}
    v_{ff} = 2930 \cdot \left( \frac{M_{1}+M_{2}}{10^{15}M_{\odot}} \right)^{1/2} \cdot \left( \frac{d}{1~\text{Mpc}} \right)^{-1/2} \\
    \cdot \left[ \frac{1-d/d_{0}}{1-(b/d_{0})^{2}} \right]^{1/2}~\text{km~s}^{-1},
\end{multline}
where $M_{1}$ and $M_{2}$ represent the respective halo masses, $b$ is the impact parameter evaluated at the initial separation $d$, and $d_0$ is the theoretical maximum separation where the two halos are assumed to be at rest before their initial gravitational collapse. 
We define a cut radius $r_{cut} = d + \max[r_{200,1}, r_{200,2}]$ so that the two halos' density profiles join smoothly along the merger axis at the initial separation, without truncation gaps that could distort the early gas dynamics.
We adopt adaptive gravitational softening for all particle types, with a minimum softening length of 1 kpc and 32 neighbors for the kernel estimation, ensuring that the force resolution scales with the local particle density \citep{Hopkins2015}.
Each simulation uses $2\times10^5$ dark matter and $4\times10^5$ gas particles for the lower-mass systems ($M_2 \leq 10^{14}\,M_\odot$), and $2\times10^6$ dark matter and $4\times10^6$ gas particles for the higher-mass systems ($M_2 \geq 5\times10^{14}\,M_\odot$).

Our simulation framework explores an $N$-dimensional grid of physical parameters ($N=6$) to enable statistical comparison with observed radio relic clusters. These parameters include the gas profile slope ($\beta$), the masses of the primary and secondary halos ($M_1, M_2$), the collision velocity factor ($f_v$), the impact parameter ($b$), and the self-interaction cross-section ($\sigma/m$). By varying these inputs, each simulation run produces a comprehensive time series tracking the time since collision ($t_{sc}$, abbreviated TSC in figures), projected halo distance ($d_h$), and projected shock distance ($d_s$) across a wide range of viewing angles ($\alpha$). 

The full range of values explored in our fiducial grid, which forms the basis for our subsequent likelihood sampling and parameter estimation, is summarized in Table \ref{tab:sim_grid}.
To represent the diverse mass scales observed in our gold sample, we initialize the secondary halo mass, $M_{2}$, at four distinct scales: $10^{13}$, $10^{14}$, $5 \times 10^{14}$, and $10^{15}~M_{\odot}$. For each scale, we define the primary halo mass, $M_{1}$, using the mass ratio $f_{M} = M_{1}/M_{2}$, with values of $f_{M} = 1$ and 3. This configuration effectively covers mass pairings in major mergers ($f_{M}\leq 3$) ranging from low-mass group-scale mergers such as PSZ2~G181.06+48.47 \citep[e.g.,][]{Ahn2025} to extreme, high-mass systems such as the ACT-CL~J0102-4915 cluster \citep[``El Gordo",][]{Kim2021}.
The maximum impact parameter at each mass scale is set according to $b_{\rm max} = 500\,(M_2/10^{14}\,M_\odot)^{\log_{10}2}$ kpc, approximately half of $r_{200}$ of the smaller halo, ensuring that the simulated range of off-axis collisions remains physically relevant across the grid.

\subsection{Determination of halo and shock positions}\label{sec:halo_shock}
To locate the halo centers in our simulations, we employ the shrinking sphere method \citep{Power2003}, which is widely used to find density peaks in $N$-body simulations. We iteratively calculate the mean position of dark matter particles within a radius that is reduced by a factor $f$ in each step until a threshold radius is reached. Halo membership is determined based on particle IDs to measure the center positions of both groups separately. 

For the shock positions, we utilize the velocity divergence $\nabla \cdot v$, as gas particles reach maximum compression at the shock front where $\nabla \cdot v$ is most negative. We examine the one-dimensional $\nabla \cdot v$ profile along the merger axis in two-dimensional density-weighted projection maps and locate the two shock positions with the minimum velocity divergence.

\begin{deluxetable}{cccccc}
\tablecaption{Simulation Parameter Grid \label{tab:sim_grid}}
\tablecolumns{6}
\tablewidth{0pt}
\tablehead{
\colhead{$M_2$ [$10^{14}M_\odot$]} & \colhead{$f_M~(M_{1}/M_{2})$} & \colhead{$f_v$} & 
\colhead{$\beta$} & \colhead{$b$ [kpc]} & \colhead{$\sigma/m$ [cm$^2$\,g$^{-1}$]}
}
\startdata
0.1 & 1, 3 & 0.4, 0.8, 1.2 & 0.57, 0.67, 0.77 & 0, 250  & 0, 0.5, 1, 3 \\
1   & 1, 3 & 0.4, 0.8, 1.2 & 0.57, 0.67, 0.77 & 0, 500  & 0, 0.5, 1, 3 \\
5   & 1, 3 & 0.4, 0.8, 1.2 & 0.57, 0.67, 0.77 & 0, 814  & 0, 0.5, 1, 3 \\
10  & 1, 3 & 0.4, 0.8, 1.2 & 0.57, 0.67, 0.77 & 0, 1000 & 0, 0.5, 1, 3 \\
\enddata
\tablecomments{Summary of the $N$-dimensional grid of physical parameters used to initialize the hydrodynamic merger simulations. The predicted shock-to-shock and halo-to-halo distances are derived via interpolation across these 576 fiducial simulation runs: piecewise quadratic in $\log_{10} M_2$ and linear in all other dimensions (see \S\ref{sec:interpolation}).}
\end{deluxetable}

\subsection{Geometric Projection Effects}\label{sec:projection}
The observable parameters in our analysis, such as the separation between cluster halos and the distance between radio-traced shock fronts, are inherently subject to geometric projection along the line of sight. To map our three-dimensional (3D) simulation data to the two-dimensional (2D) plane of the sky, we must account for the viewing angle ($\alpha$), defined as the angle between the merger axis and the plane of the sky. 

\begin{figure}
\centering
\includegraphics[width=8.5cm]{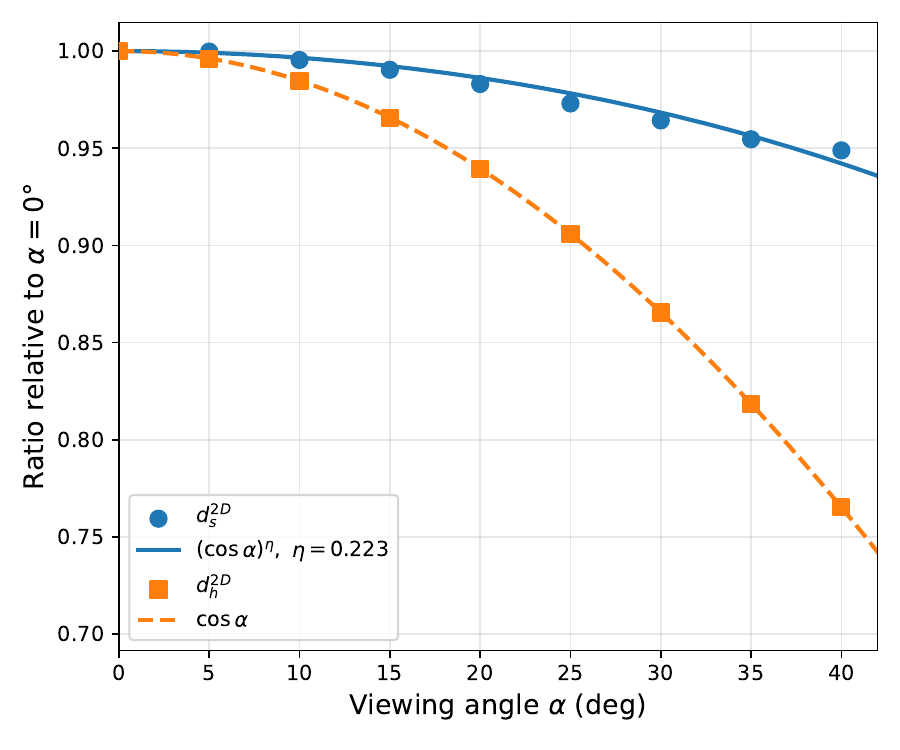}
\caption{Projected shock-to-shock and halo-to-halo distances as a function of viewing angle. We illustrate a head-on collision case where $M_1=M_2=5\times10^{14}\,M_{\sun}$, $f_v=1.2$, and $\sigma/m=0~\text{cm}^{2}\,\text{g}^{-1}$. 
Both distances are normalized to unity at $\alpha = 0\degr$.
The blue (orange) data points represent normalized projected shock-to-shock (halo-to-halo) distances measured at different viewing angles. As expected, the halo-to-halo distances perfectly scale as $\cos \alpha$ (orange line). On the other hand, the dependence on the viewing angle is much weaker in shock-to-shock distances. For instance, at $\alpha=30\degr$, the projected shock-to-shock distance decreases by only $\sim3$\%. We model this weaker dependence as $d_{s}^{2D} = \cos^{\eta} \alpha \cdot d_{s}^{3D}$; fitting our simulation ensemble yields a best-fit value of $\eta \approx 0.223$ (blue line), which provides an excellent ($\lesssim1\%$) description of this merger case.}
\label{fig:projection_effect}
\end{figure}

The projected halo-to-halo distance is straightforward and can be modeled following a standard trigonometric projection:
\begin{equation}
d_{h}^{2D} = \cos \alpha \cdot d_{h}^{3D}.
\end{equation}
In contrast, because the radio-emitting shock fronts are highly extended and parts of quasi-spherical surfaces, their projected positions are much less sensitive to the viewing angle than the halo separation.
Figure \ref{fig:projection_effect} illustrates a head-on collision case where $M_1=M_2=5\times10^{14}\,M_{\sun}$, $f_v=1.2$, and $\sigma/m=0~\text{cm}^{2}\,\text{g}^{-1}$.  
We model this deviation using a power-law modification:
\begin{equation}
d_{s}^{2D} = \cos^{\eta} \alpha \cdot d_{s}^{3D}.
\end{equation}
By fitting the equation to our simulation ensemble across a range of viewing angles, we find a best-fit value of $\eta \approx 0.223$, which leads to an accuracy of $\lesssim1\%$.

\subsection{Interpolation Scheme for Off-Grid Mergers}\label{sec:interpolation}

Our fiducial simulation grid is discrete by construction, yet the MCMC
sampler requires the predicted distances $d_s$ and $d_h$ to be evaluated
at arbitrary parameter values. We therefore build a continuous forward model
by interpolating across the 576 simulation outputs.

The forward model is a seven-dimensional interpolator with axes
$(\beta,\, M_2,\, f_M,\, f_v,\, \hat{b},\, \sigma/m,\, t_{sc})$, where
$f_M = M_1/M_2 \geq 1$ is the mass ratio and $\hat{b} = b/b_{\rm max}(M_2)$ is
the impact parameter normalized by the maximum grid value at the mass scale of
the smaller halo (Table~\ref{tab:sim_grid}). Separate interpolators are constructed for the
predicted shock-to-shock distance ($d_s$) and halo-to-halo distance ($d_h$).

Because the secondary halo mass grid, $M_2 \in \{0.1, 1, 5, 10\}\times
10^{14}~M_{\odot}$, is strongly non-uniform in linear space, we adopt a
hybrid interpolation scheme. In the $M_2$ dimension, we apply piecewise
quadratic interpolation in $\log_{10} M_2$ space; in all remaining six
dimensions, we apply standard linear interpolation using SciPy's
\texttt{RegularGridInterpolator}\footnote{\url{https://docs.scipy.org/doc/scipy/reference/generated/scipy.interpolate.RegularGridInterpolator.html}}. Specifically, for each query point, we
evaluate the six-dimensional linear interpolator at every $M_2$ grid node
and then fit a piecewise quadratic through the resulting values as a function
of $\log_{10} M_2$ to obtain the final prediction. The projection corrections
derived in \S\ref{sec:projection} are then applied: $d_s^{2D} = \cos^{0.223}\alpha \cdot
d_s^{3D}$ and $d_h^{2D} = \cos\alpha \cdot d_h^{3D}$.

To validate the interpolation scheme, we generate 20 sets of randomized merger parameters drawn from within the grid bounds but offset from any grid node, and run dedicated GIZMO simulations for each. Comparing the interpolated predictions to the directly simulated shock-to-shock and halo-to-halo distances, we find agreement well within 2\% in all cases. This confirms that the hybrid quadratic--linear interpolation captures the nonlinear dependence on $M_2$ with sufficient accuracy for our Bayesian inference. Although this 2\% accuracy is, in principle, a model-side systematic, it is subdominant to the observational uncertainties on the halo-to-halo distance (typically $\sim 10$--$30\%$ at weak-lensing precision; Table~\ref{tab:gold}), which dominate the per-system likelihood width. The propagated effect on the joint $\sigma/m$ posterior is therefore negligible.

\subsection{Sensitivity of Halo and Shock Distances to SICS}\label{sec:sensitivity}

A critical finding from our simulations is the distinct sensitivity of halo and shock distances to the SICS (Figure~\ref{fig:t_vs_d}). While the shock propagation speed remains nearly independent of the cross-section, the post-collision orbital velocity of dark matter halos is sensitive to the SICS. After pericenter passage, as the cross-section increases, the momentum-transfer drag slows the halo orbital velocity, causing $d_{h}$ to decrease while $d_{s}$ remains virtually unchanged. This decoupling allows us to use $d_s$ as a near-independent reference scale. By jointly fitting the observed $d_h$ and $d_s$ against our simulation predictions, we can isolate the SICS-dependent suppression of the halo separation.

This approach was first presented in \citet{Park2022}. \citet{Fischer2023} subsequently noted that combining shock-front positions with galaxy/BCG positions holds promise as a probe of dark matter physics, but concluded that a more detailed theoretical study would be needed before it could be turned into a quantitative constraint. The present work advances this idea into a rigorous quantitative constraint through a Bayesian framework calibrated against 576 simulations and applied to eleven gold cluster mergers.

\subsection{Likelihood Sampling and Parameter Estimation}\label{sec:likelihood}

To derive constraints on the dark matter self-interaction cross-section ($\sigma/m$), we employ a Bayesian inference framework. We explore the multidimensional parameter space by sampling the posterior distribution, which is proportional to the product of our likelihood function and the prior distributions of our model parameters.

For the sampling process, we utilize the \texttt{emcee} Python package \citep{ForemanMackey2013}, which implements an affine-invariant ensemble Markov Chain Monte Carlo (MCMC) sampler. 
The posterior probability distribution is obtained by sampling an 8-dimensional parameter space $\boldsymbol{\theta} = (\beta, M_{1}, M_{2}, f_{v}, b, t_{sc}, \alpha, \sigma/m)$, where $M_{1}$ and $M_{2}$ are the masses of the two subclusters, $f_{v}$ is the collision velocity factor, $\beta$ is the gas profile index, $b$ is the impact parameter, $\sigma/m$ is the SICS, $t_{sc}$ is the time-since-collision, and $\alpha$ is the viewing angle. For any given parameter set $\boldsymbol{\theta}$, we employ a function $S$ that maps these parameters to the predicted projected shock-to-shock and halo-to-halo distances, $\mathbf{d}^{p}=[d_{s}^{p}, d_{h}^{p}]$. 
The function $S$ is the interpolation scheme described in \S\ref{sec:interpolation}.

The log-likelihood for a single cluster system is defined as:
\begin{equation}
    \ln(\text{likelihood}) = -0.5 \times \sum \left( \frac{\mathbf{d}^{o} - \mathbf{d}^{p}}{\boldsymbol{\sigma}^{o}} \right)^{2}
\end{equation}
where $\mathbf{d}^{o} = [d_{s}^{o}, d_{h}^{o}]^{T}$ represents the observed shock-to-shock and halo-to-halo distances, and $\boldsymbol{\sigma}^{o} = [\sigma_{d_{s}}, \sigma_{d_{h}}]^{T}$ represents their corresponding measurement uncertainties. 

We specify prior distributions that reflect both the physical constraints of cluster merger dynamics and the observational uncertainties of the gold sample. We adopt log-normal priors for the halo masses $M_{1}$ and $M_{2}$, centered on the weak-lensing estimates provided in Table~\ref{tab:gold}. The use of log-normal priors is physically motivated by the positive-definite nature of mass and the characteristic uncertainty distributions of lensing measurements.

For the remaining nuisance parameters, we employ flat priors over physically motivated ranges to ensure that our constraints on $\sigma/m$ are not biased by overly restrictive assumptions:
\begin{itemize}
\item \textbf{Gas Profile ($\beta$):} A flat prior is set over the range $[0.57, 0.77]$, which spans the values typically obtained when fitting single $\beta$-models to the X-ray surface brightness profiles of massive, relaxed galaxy clusters \citep[e.g.,][]{Jones1984, Mohr1999}.
\item \textbf{Collision Velocity ($f_{v}$):} The collision velocity factor $f_v$ is assigned a flat prior over $[0.7, 1.0]$. Since $f_v=1$ corresponds to the free-fall speed (the theoretical maximum for a two-body system infalling from rest at infinity), values slightly below unity are typical for cosmological infall. Cosmological simulations show that the pairwise infall velocities of massive halos at first virial crossing are distributed near $f_v \sim 0.9$ with a dispersion of $\sim$0.2 \citep{Benson2005, Wetzel2011}, so $[0.7, 1.0]$ encompasses the bulk of the distribution around the theoretical free-fall maximum.
\item \textbf{Viewing Angle ($\alpha$):} We assume a flat viewing angle prior $[0^\circ, 40^\circ]$. Double radio relics are subject to a strong projection bias: the limb-brightening of the quasi-spherical shock front dramatically increases the radio surface brightness when the merger axis lies near the plane of the sky, making such systems the most readily identified double-relic population in flux-limited surveys \citep{vanWeeren2019}. Accordingly, we restrict $\alpha$ to $<40^\circ$, beyond which the probability of identifying a system as a double relic drops sharply.
\item \textbf{Impact Parameter ($b$):} A flat prior is adopted for $b$ over $[0,\, 0.2\,b_{\rm max}]$, where $b_{\rm max}$ is the maximum impact parameter in our simulation grid at the relevant mass scale (Table~\ref{tab:sim_grid}). This restricts the analysis to nearly head-on collisions, consistent with the finding from hydrodynamical simulations that the symmetric shocks necessary to produce double radio relics form preferentially for $b \lesssim 0.2\,R_{\rm vir}$ \citep{Kang2012, Ha2018}; we adopt a more conservative bound to ensure clean, well-separated double-relic morphologies.

\item \textbf{Merger Phase ($t_{sc}$):} We adopt a flat prior for $t_{sc}$ of $[0.4, 1.0]$ Gyr. This window targets the active phase of radio relic evolution. Hydrodynamical simulations of binary mergers show that the kinetic energy flux dissipated by merger shocks rises sharply $\sim 0.4$ Gyr after core passage and fades significantly beyond $\sim 1.0$ Gyr as the shocks reach the low-density outskirts \citep{Ha2018, Lee2026}.
\item \textbf{Self-interaction Cross-section ($\sigma/m$):} A flat prior is adopted over $[0,\, 2]~\text{cm}^{2}\,\text{g}^{-1}$, spanning from the collisionless case to well above the upper range probed by our simulation grid.
\end{itemize}

Rather than computing a global likelihood for the entire sample, we estimate the individual posterior $P(\sigma/m \mid \mu_i)$ for each merger $\mu_i$ and multiply them to obtain the joint posterior. This is justified by the independence of all nuisance parameters across systems: only $\sigma/m$ is shared. A joint fit would otherwise require sampling an $N = 7 \times 11 + 1 = 78$ dimensional parameter space, which is computationally prohibitive. By marginalizing over the system-specific parameters first, we effectively harness the collective statistical power of the gold sample:
\begin{equation}
    P(\sigma/m \mid \{\mu_i\}) \propto \prod_{i=1}^{11} P(\sigma/m \mid \mu_i).
\end{equation}

\begin{deluxetable*}{lcccccc}
\tablecaption{Observed Parameters of the Eleven Gold Radio Relic Clusters \label{tab:gold}}
\tablecolumns{6}
\tablewidth{0pt}
\tablehead{
\colhead{Name} & \colhead{$z$} & \colhead{$d_{\text{shock}}$ [Mpc]} & \colhead{$d_{\text{halo}}$ [Mpc]} & \colhead{$M_1$ [$10^{14}$ M$_\odot$]} & \colhead{$M_2$ [$10^{14}$ M$_\odot$]}
}
\startdata
Abell 3376$^{a}$         & 0.046 & $2.08 \pm 0.02$ & $1.10 \pm 0.11$ & $3.00 \pm 1.50$ & $0.90 \pm 0.65$ \\
ZWCL 0008.8+5215$^{b}$   & 0.104 & $1.61 \pm 0.02$ & $1.02 \pm 0.10$ & $2.90 \pm 0.80$ & $2.70 \pm 0.80$ \\
Abell 2345$^{b}$         & 0.176 & $2.28 \pm 0.02$ & $1.00 \pm 0.10$ & $4.70 \pm 1.20$ & $3.20 \pm 1.00$ \\
CIZA J2242.8+5301$^{c}$  & 0.192 & $2.56 \pm 0.03$ & $1.07 \pm 0.29$ & $11.0 \pm 3.45$ & $9.80 \pm 3.15$ \\
Abell 1240$^{d}$         & 0.195 & $2.23 \pm 0.02$ & $1.10 \pm 0.16$ & $2.61 \pm 0.56$ & $1.09 \pm 0.39$ \\
PSZ2 G181.06+48.47$^{e}$ & 0.234 & $2.70 \pm 0.02$ & $0.50 \pm 0.12$ & $2.71 \pm 0.50$ & $0.88 \pm 0.33$ \\
RXC J1314.4-2515$^{b}$   & 0.247 & $1.50 \pm 0.01$ & $0.54 \pm 0.09$ & $4.20 \pm 1.30$ & $2.30 \pm 1.00$ \\
ZWCL 1856.8+6616$^{b}$   & 0.304 & $1.92 \pm 0.02$ & $0.66 \pm 0.09$ & $1.60 \pm 0.80$ & $1.50 \pm 0.70$ \\
MACS J1752.0+4440$^{b}$  & 0.366 & $2.10 \pm 0.02$ & $1.19 \pm 0.07$ & $5.60 \pm 1.80$ & $5.60 \pm 1.70$ \\
ZWCL1447.2+2619$^{f}$    & 0.372 & $2.01 \pm 0.02$ & $0.36 \pm 0.15$ & $2.70 \pm 0.80$ & $1.00 \pm 0.50$ \\
ACT-CL J0102-4915$^{g}$  & 0.870 & $1.82 \pm 0.02$ & $0.77 \pm 0.13$ & $9.90 \pm 2.15$ & $6.50 \pm 1.65$ \\
\enddata
\tablecomments{Observational properties of the eleven gold clusters. Asymmetric mass errors from the literature were symmetrized using arithmetic averaging to maintain central values. References: $^{a}$\cite{Monteiro-Oliveira2017}; $^{b}$\cite{Finner2025}; $^{c}$\cite{Jee2015}; $^{d}$\cite{Cho2022}; $^{e}$\cite{Ahn2025}; $^{f}$\cite{Lee2022}; $^{g}$\cite{Kim2021}.}
\end{deluxetable*}

\begin{figure}
\includegraphics[width=8.5cm]{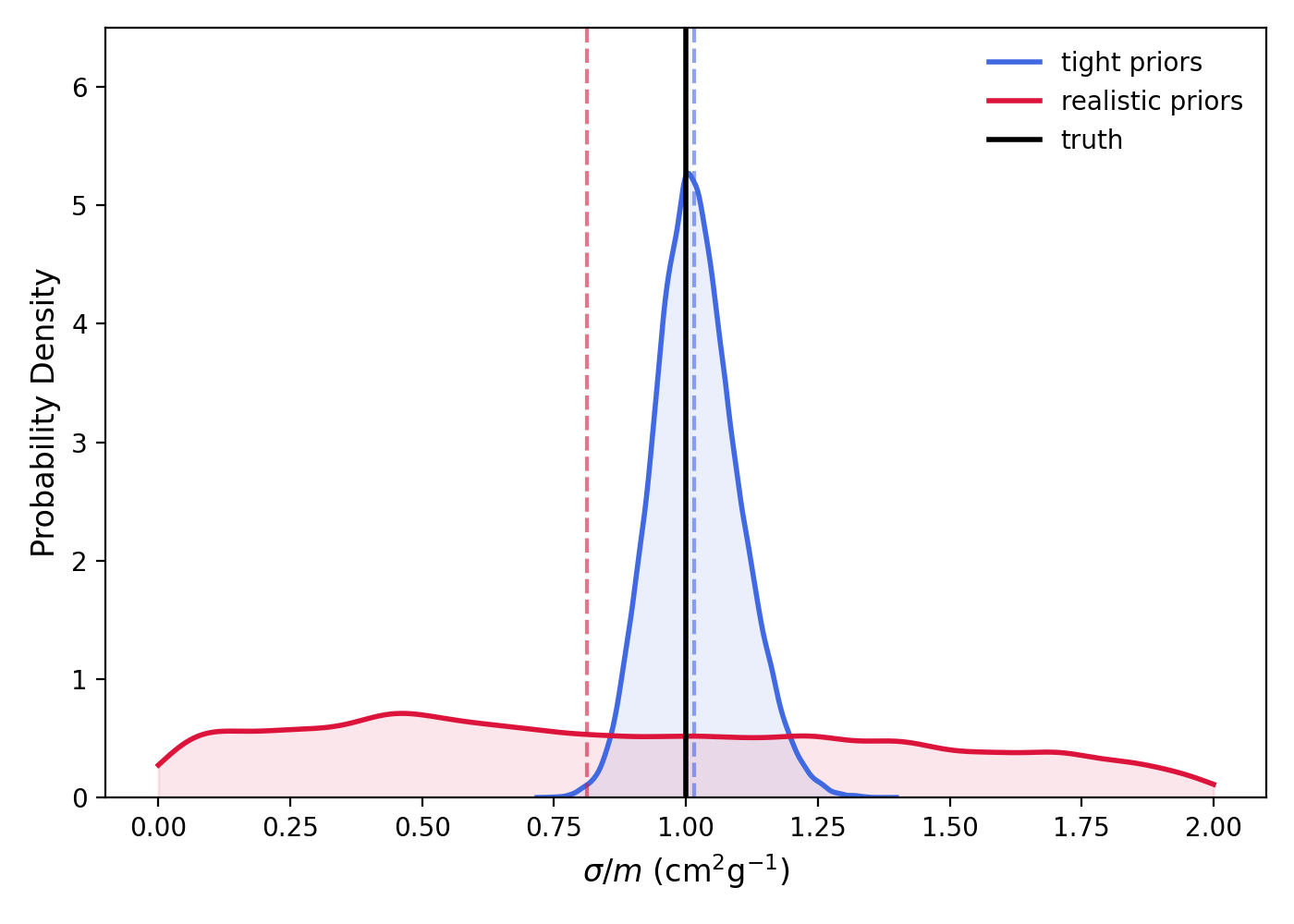}

\caption{SICS recovery test with a single mock radio relic cluster. Solid curves represent posteriors while the vertical dashed lines indicate the medians.
When we employ tight priors centered on the true physical parameters of the merger, the resulting posterior distribution (blue) peaks sharply at the true $\sigma/m=1$ value. Conversely, when the priors are relaxed to reflect a realistic observational scenario, the constraining power (red) is substantially diminished.
}
\label{fig:sigma_m_single}
\end{figure}

\begin{figure*}
\includegraphics[width=8.0cm]{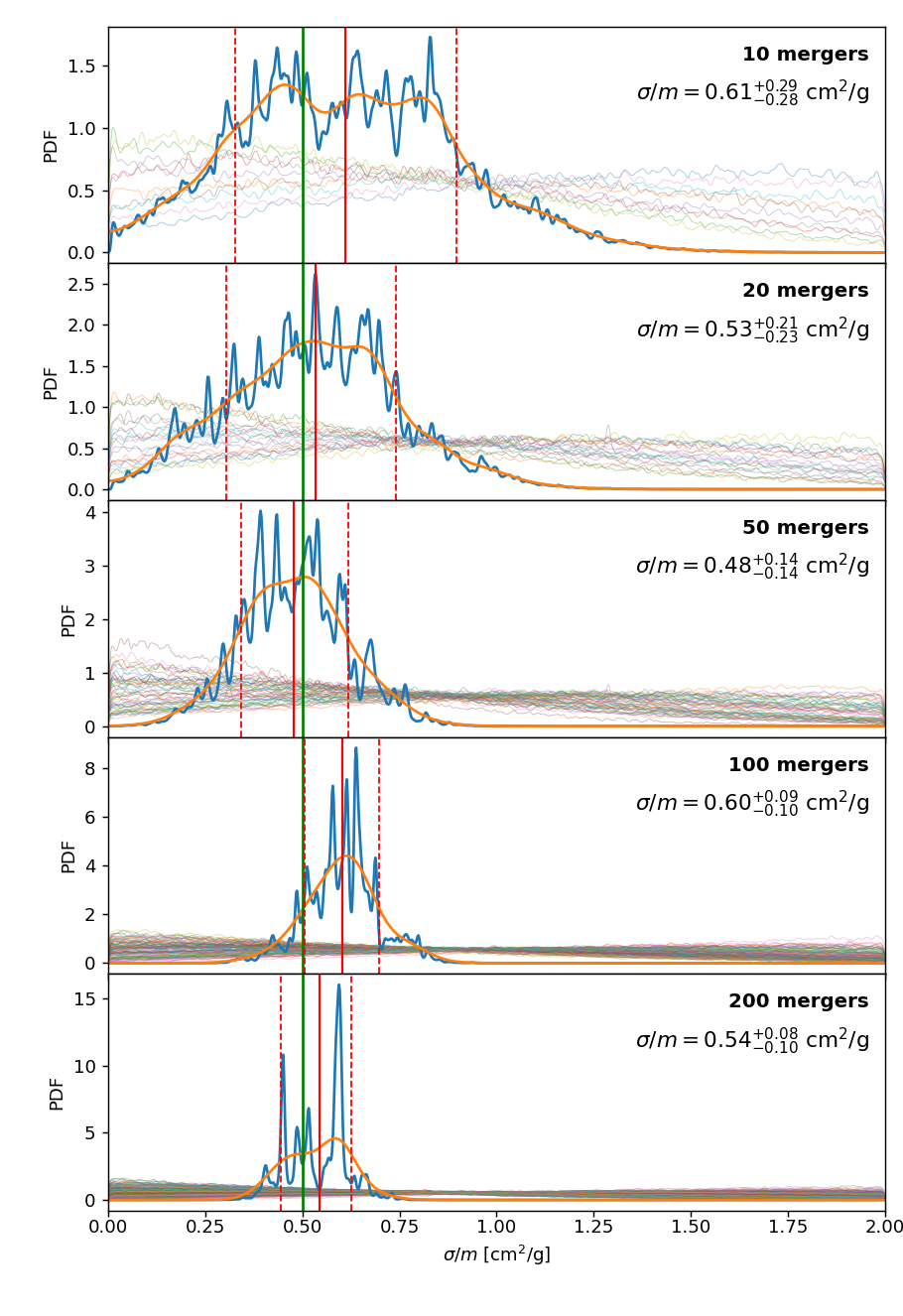}
\includegraphics[width=8.0cm]{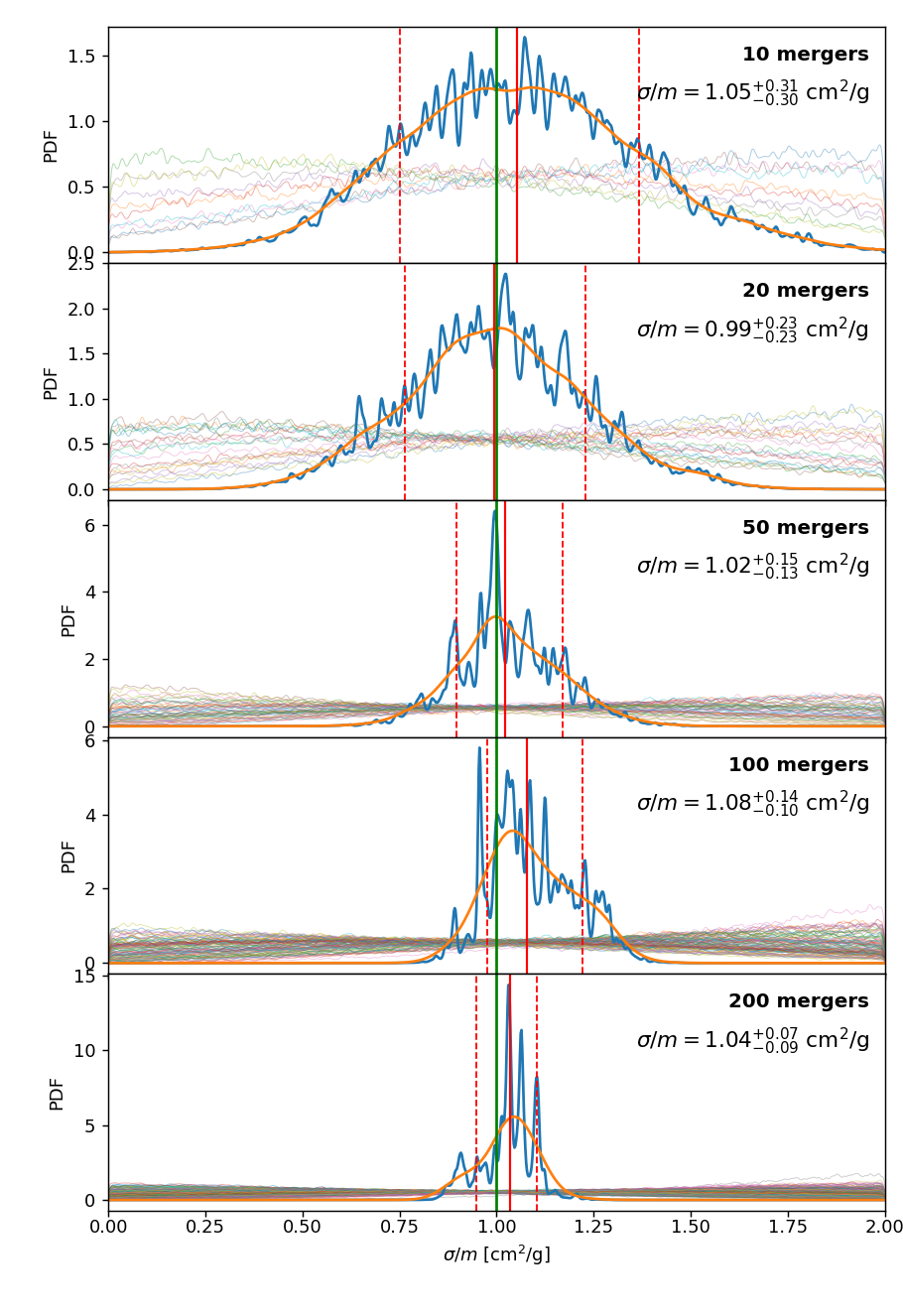}

\caption{SICS recovery test with multiple mergers.
\editcap{The true cross-sections are $\sigma/m=0.5~\text{cm}^{2}\,\text{g}^{-1}$ and $1~\text{cm}^{2}\,\text{g}^{-1}$ in the left and right panels, respectively, as indicated by the green vertical lines.}
The faint thin lines indicate the individual posteriors $p(\sigma/m \mid \mu_i)$ derived from each $i^{th}$ merger data $\mu_i$. The blue solid line represents the (unsmoothed) combined posterior distribution $p(\sigma/m \mid \{\mu\}) \propto \prod_{i=1}^{N} p(\sigma/m \mid \mu_i)$. The orange line represents the smoothed Kernel Density Estimate (KDE) of this combined posterior. The red vertical solid (dashed) line shows the median ($\pm 1\sigma$ limit). 
As the number of mergers $N$ increases, the joint posterior distribution narrows with its error shrinking as $\sim 1/\sqrt{N}$.
}
\label{fig:sigma_m_multi}
\end{figure*}

\subsection{Methodological Validation with Mock Data}\label{sec:validation}

To demonstrate the reliability of our statistical framework before applying it to observed clusters, we performed two sets of validation tests using mock data generated from our simulation grid. First, we assess how well the method recovers a known input cross-section for a single mock merger under idealized and realistic prior assumptions (\S\ref{sec:single_merger}). Second, we demonstrate how the constraining power improves as the number of independent mergers increases (\S\ref{sec:multi_merger}).

\subsubsection{Single Merger Case Study}\label{sec:single_merger}

We first evaluate the sensitivity of our method using a single mock merger event. The physical parameters for this test are configured as follows: $\beta=0.67$, $M_{1}=9.9\times10^{14}\,M_{\sun}$, $M_{2}=6.5\times10^{14}\,M_{\sun}$, $f_{v}=0.8$, $b=200~\text{kpc}$, $t_{sc}=0.75~\text{Gyr}$, and $\alpha=30\degr$, assuming a true cross-section of $\sigma/m=1~\text{cm}^{2}\,\text{g}^{-1}$. The mass values for this case study are adopted from the ``El Gordo'' cluster \citep{Kim2021}.

We investigate the impact of prior distributions on the recovered SIDM cross-section by considering two distinct scenarios for the seven nuisance parameters ($\beta$, $M_{1}$, $M_{2}$, $f_{v}$, $b$, $t_{sc}$, and $\alpha$):

\begin{enumerate}
    \item \textbf{Idealized Scenario (Tight Priors):} We adopt unrealistically narrow flat priors with widths of 1\% of the true value on all nuisance parameters, simulating near-perfect knowledge of the merger's geometry and physical properties except for $\sigma/m$. Under these idealized conditions, the resulting posterior distribution for $\sigma/m$ is sharply peaked at the true input value (blue curve in Figure~\ref{fig:sigma_m_single}). This confirms that the ratio of halo-to-shock distances serves as a high-fidelity indicator of dark matter self-interaction when the merger phase and viewing angle are precisely constrained.
    
    \item \textbf{Realistic Scenario (Broad Priors):} We impose broader priors that reflect typical observational uncertainties and merger ambiguities. Specifically, we adopt the prior ranges described in \S\ref{sec:likelihood}. For the masses, we use log-normal priors with medians and linear-scale dispersions of $M_{1} = (9.90 \pm 2.15) \times 10^{14}~M_{\odot}$ and $M_{2} = (6.50 \pm 1.65) \times 10^{14}~M_{\odot}$ \footnote{We adopt the weak-lensing mass measurements and uncertainties from \cite{Kim2021}.}. As shown in Figure~\ref{fig:sigma_m_single}, the constraining power for a single merger is substantially diminished (red curve). Marginalization over these degenerate parameters significantly broadens the posterior distribution, illustrating why single-cluster studies are generally insufficient for providing precise dark matter constraints.
\end{enumerate}

\subsubsection{Improving Constraining Power with Multiple Mergers}\label{sec:multi_merger}

To overcome the weak constraining power of individual mergers when realistic priors are applied, we demonstrate the efficacy of our stacking methodology by increasing the sample size. Using the same realistic priors as in \S\ref{sec:single_merger}, we generated mock observational data sets for \edit1{two true cross-sections of $\sigma/m = 0.5$ and $1~\text{cm}^{2}\,\text{g}^{-1}$} and analyzed the resulting joint posteriors for samples of 10, 20, 50, 100, and 200 mergers.

As illustrated in Figure~\ref{fig:sigma_m_multi}, the posterior distribution $P(\sigma/m)$ becomes increasingly sharper as the sample size grows. \edit1{In both cases, the posteriors peak near the respective true values and the true $\sigma/m$ is recovered within the 68\% credible interval.}

While a single cluster collision provides only marginal information, the accumulation of independent data points effectively averages out the merger-to-merger variations in nuisance parameters such as merger phase and viewing angle. This validation confirms that our method can achieve high precision by combining multiple systems, and motivates our strategy of applying the joint analysis to the eleven gold mergers in \S\ref{sec:results}.

\section{Results}\label{sec:results}

Having validated the statistical framework with mock data (\S\ref{sec:validation}), we now apply the method to our gold sample of observed double radio relic clusters. We describe the sample selection criteria in \S\ref{sec:sample_selection} and present the resulting SICS constraint in \S\ref{sec:constraints}.

\begin{figure}
\includegraphics[width=9cm]{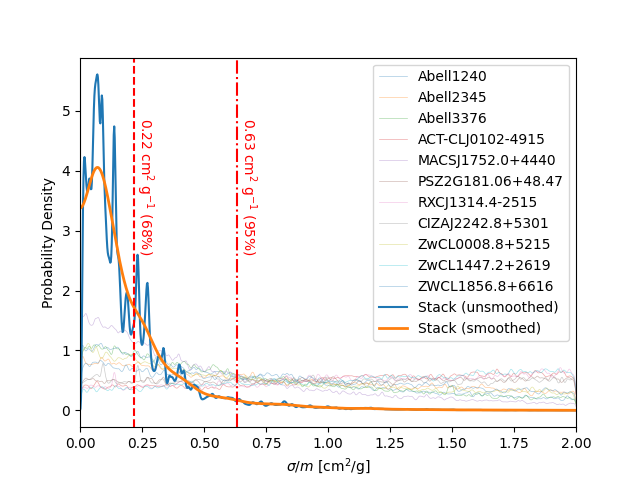}

\caption{Constraints on the SICS from real radio relic clusters. The faint thin lines indicate the individual posteriors $p(\sigma/m \mid \mu_i)$ derived from each merger $\mu_i$. The blue solid line is the combined posterior and the orange line its smoothed KDE. The red vertical dashed and dot-dashed lines mark the 68\% and 95\% upper limits, respectively.
}
\label{fig:sigma_m_real}
\end{figure}

\subsection{Sample Selection}\label{sec:sample_selection}
With the increased sensitivity of new-generation radio surveys, the census of known radio relic clusters has grown to approximately 100--120 \citep[e.g.,][]{vanWeeren2019, Knowles2022}. For this study, we defined a gold sample of 11 targets selected based on the following criteria:

\begin{enumerate}
    \item \textbf{Bimodality:} The system displays a clear bimodal morphology in both galaxy density and weak lensing (WL) mass maps.
    \item \textbf{Mass Precision:} High-fidelity WL mass estimates are available for both subclusters.
    \item \textbf{Radio Morphology:} The system hosts double radio relics.
    \item \textbf{Geometry:} The relics are symmetric, and the relic--relic vector is well-aligned with the merger axis inferred from the mass distribution (the mass--mass vector) \citep{Lee2026}.
\end{enumerate}
\noindent
Table~\ref{tab:gold} summarizes the observational properties of the 11 gold clusters, which serve as constraints for our MCMC inference. 

\subsection{Constraints on $\sigma/m$}\label{sec:constraints}

Applying this methodology to the eleven gold clusters, we obtain individual marginalized posteriors and combine them into a joint posterior. Each individual posterior encodes the full degeneracy between $\sigma/m$ and the system-specific nuisance parameters, which are then integrated out, leaving only the SICS dependence to propagate into the joint constraint.

As shown in Figure~\ref{fig:sigma_m_real}, the combined analysis yields a 68\% (95\%) upper limit of $\sigma/m < 0.22~\text{cm}^{2}\,\text{g}^{-1}$ ($< 0.63~\text{cm}^{2}\,\text{g}^{-1}$) on the dark matter self-interaction cross-section. This is the most stringent constraint on the SICS derived from cluster-scale collisions to date. The constraining power arises from the combination of eleven independent merger systems, each contributing an individual posterior that is marginalized over seven nuisance parameters: halo masses, gas profile slope, collision velocity, impact parameter, merger phase, and viewing angle. The joint posterior is well-peaked and smoothly declining, with no significant support at large cross-sections, indicating a robust preference for low or vanishing self-interaction at cluster scales. Systematic effects and the sensitivity of this result to prior choices are examined in \S\ref{sec:discussions}.

\section{Discussion}\label{sec:discussions}
\begin{figure*}
\centering
\includegraphics[width=\textwidth]{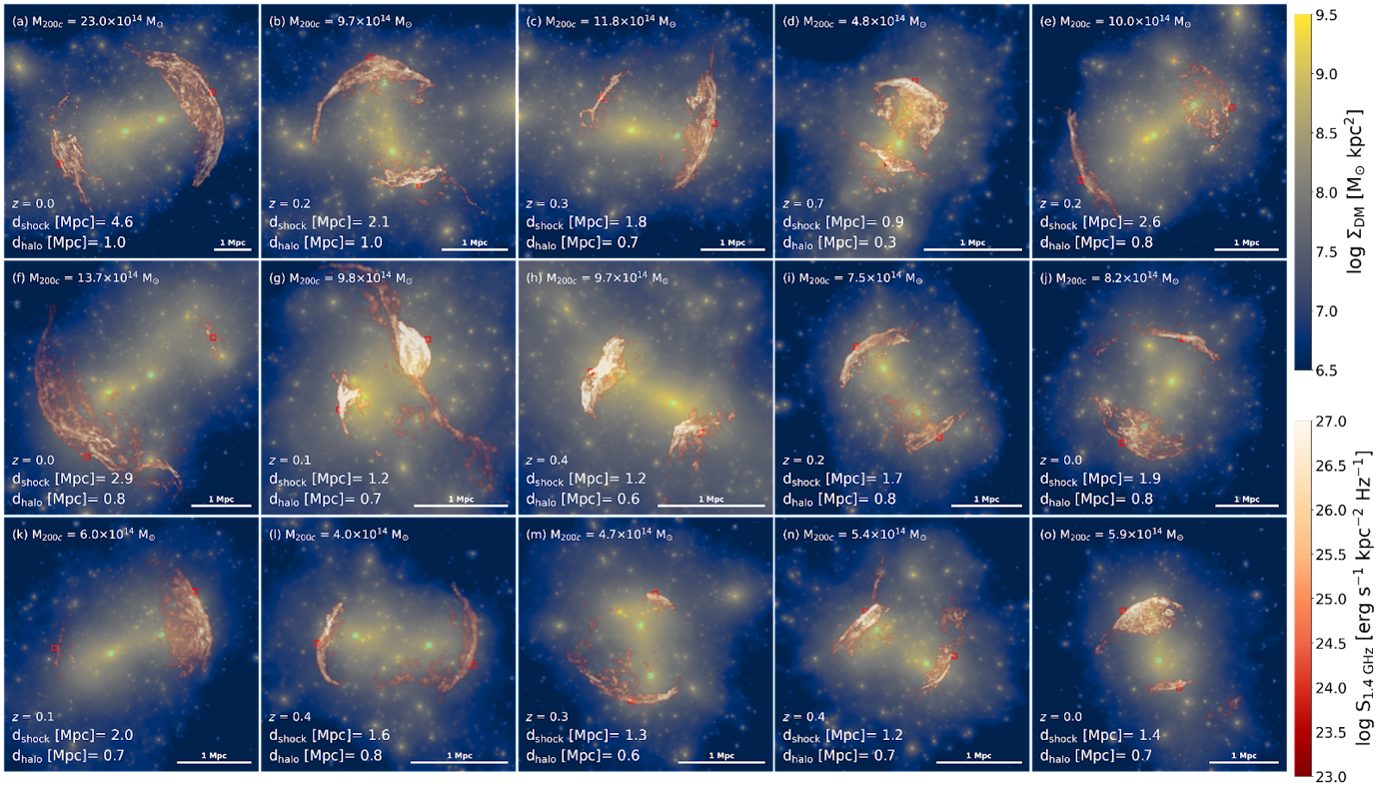}
\caption{Radio relic galaxy cluster analogs from the TNG-Cluster simulation used for validation.
Analogous to our observational gold sample selection, we identify 15 analogs within the TNG-Cluster simulation that experience major binary mergers with small impact parameters and clear symmetric double radio relics.
These systems are selected as massive halo pairs ($M_{\text{halo}} \ge 10^{14}\,M_{\odot}$) undergoing active, first-passage collisions. The background images display the projected dark matter density distribution, with green circles indicating the halo centers. Red contours represent projected radio emissions, which trace the locations of merger shocks. Red rectangles denote the identified shock front positions.}
\label{fig:tngcluster_sample}
\end{figure*}

\begin{figure}
\includegraphics[width=8.5cm]{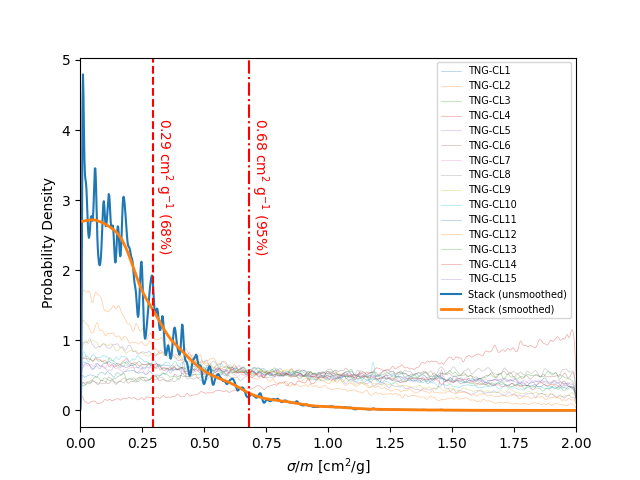}
\caption{Same as Figure~\ref{fig:sigma_m_real}, but based on synthetic data from the 15 TNG-Cluster analogs shown in Figure~\ref{fig:tngcluster_sample}. The TNG-Cluster simulation models dark matter with no self-interaction ($\sigma/m=0~\text{cm}^{2}\,\text{g}^{-1}$). We constrain the SICS upper limit to a level comparable to the observational result.
}
\label{fig:tngcluster_sics}
\end{figure}

\begin{figure}
\includegraphics[width=8.5cm]{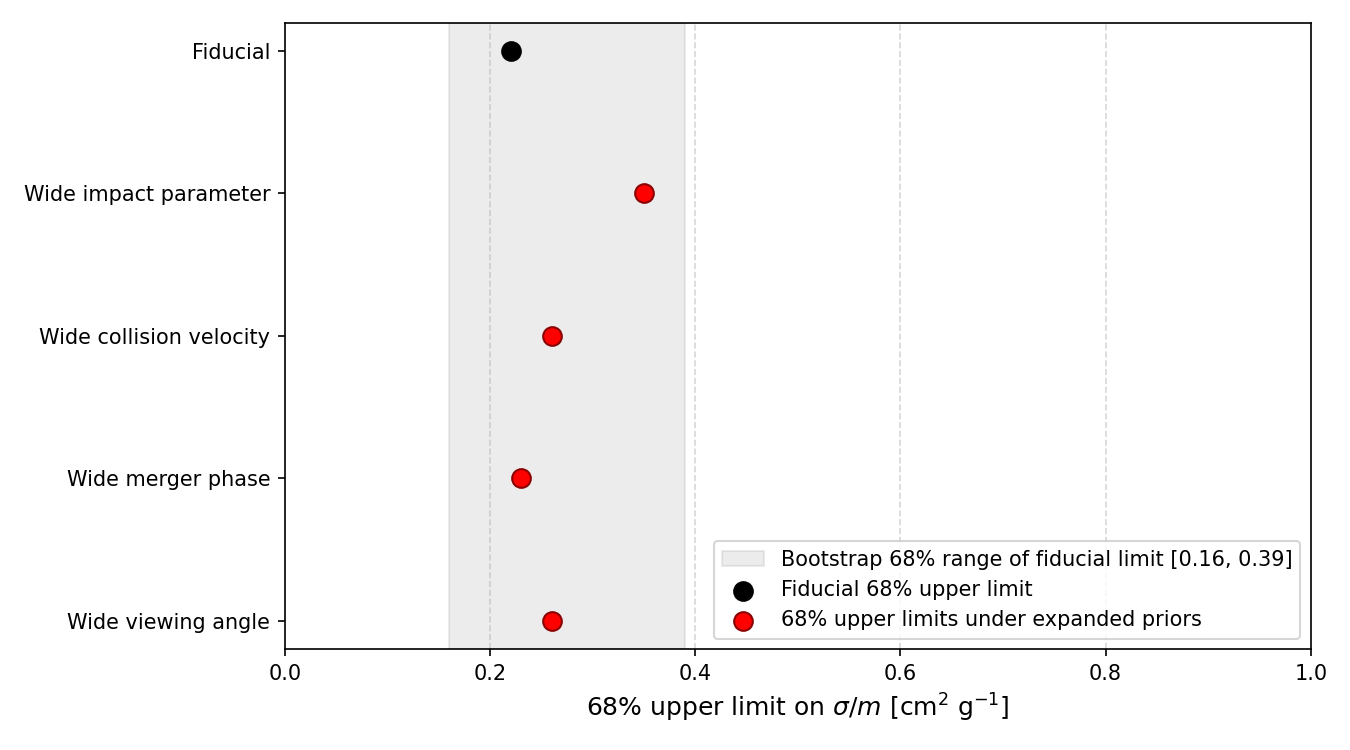}
\caption{Sensitivity of the $\sigma/m$ upper limit to prior assumptions.
\editcap{Each red point marks the joint 68\% upper limit obtained when one prior range is expanded as described in the text, and the black point shows the fiducial 68\% upper limit. The shaded band indicates the 68\% bootstrap range of the fiducial upper limit, derived from 1,000 resamples of the eleven-cluster sample. This band characterizes the sample-to-sample variability of the limit itself and is not a credible interval on $\sigma/m$. All four prior variations fall within this band.}
}
\label{fig:prior_sensitivity}
\end{figure}

\subsection{Validation with TNG-Cluster}\label{sec:tngcluster}
For our constraint on $\sigma/m$ with observations, we rely on idealized simulations, which do not capture physical complexities such as surrounding large-scale structures, halo substructures, and mass accretion. These issues are somewhat mitigated in the current study because we carefully select bimodal merger cases and ensure that their double radio relics align with the halo-to-halo vectors. 

Nevertheless, it is important to test if we can achieve similar levels of constraining power when the aforementioned physical complexities are included.
Here we utilize the cosmological zoom-in simulation TNG-Cluster \citep{Nelson2024, Lee2024, Lee2026} and demonstrate that our approach is still valid in the cosmological context.
TNG-Cluster is an extension to the IllustrisTNG suite of cosmological magnetohydrodynamical simulations.
The project re-simulates 352 cluster regions from a parent $(1\,\rm Gpc)^3$ simulation volume. Since dark matter self-interaction is not included in the TNG-Cluster simulation, the true cross-section to be recovered is $\sigma/m=0~\text{cm}^{2}\,\text{g}^{-1}$. 

Analogous to our observational gold sample selection, we identify 15 analogs within the TNG-Cluster simulation that experience major binary mergers with small impact parameters and clear symmetric double radio relics, as shown in Figure~\ref{fig:tngcluster_sample}. The selected systems have mass ratios $f_M < 3$ and smaller halo masses in the range $10^{14}\,M_\odot \lesssim M_2 \lesssim 5\times10^{14}\,M_\odot$.
Using the same prior ranges and methodology applied to our gold observational sample, we derive the SICS constraint (Figure~\ref{fig:tngcluster_sics}).
The 68\% (95\%) upper limit is $\leq 0.32~\text{cm}^{2}\,\text{g}^{-1}$ ($\leq 0.75~\text{cm}^{2}\,\text{g}^{-1}$). Given that the exact value varies with the sample selection, as expected from the small sample of 15 systems, this result is consistent within the 68\% credible interval of the observational result.

\subsection{Bias and Prior Sensitivity Analysis}\label{sec:prior_sensitivity}
While the total uncertainty budget in the current study is dominated by statistical variance from the limited gold sample size, the sensitivity of our results to prior assumptions must be assessed. As future surveys provide much larger samples, prior-driven systematics will become the limiting factor.

We investigate the systematic dependence of our $\sigma/m$ measurements by examining the following four expanded prior assumptions:
\begin{itemize}
\item \textbf{Viewing angle prior:} We increase the current prior upper bound from $\alpha \le 40\degr$ to $\alpha \le 70\degr$. This accounts for the possibility that some of our eleven gold mergers may be occurring at more extreme inclinations relative to the line of sight \citep[e.g., PSZ2 G181.06+48.47;][]{Rajpurohit2025}.
\item \textbf{Impact parameter prior:} We increase the maximum allowed impact parameter ($b$) by 50\%, allowing for more off-axis collision geometries.
\item \textbf{Collision velocity prior:} We expand the normalized infall velocity ($f_v = v_{rel}/v_{ff}$) prior range from $[0.7, 1.0]$ to $[0.4, 1.2]$. This covers a broader range of the pairwise velocity distribution, including the lower-velocity population identified in large-scale simulations.
\item \textbf{Merger phase prior:} We enlarge the TSC interval from $[0.4, 1.0]$~Gyr to $[0.1, 1.6]$~Gyr to cover a wider range of dynamical stages.
\end{itemize}

\edit1{Figure~\ref{fig:prior_sensitivity} illustrates the resulting variations in the $\sigma/m$ posterior. To judge whether these prior-induced shifts are significant, we require a yardstick for how much the upper limit itself fluctuates when the makeup of the gold sample changes. We therefore quantified this sample-to-sample variability by performing 1,000 bootstrap resamples of our eleven-cluster dataset. We stress that the resulting band measures the robustness of the recovered limit against changes in sample composition; it is \emph{not} a credible interval on $\sigma/m$, and the spread of the band should not be read as an uncertainty attached to the reported upper limit. As indicated by the shaded region and error bars in Figure~\ref{fig:prior_sensitivity}, the distribution of bootstrap limits is highly skewed. This skewness is attributed to the heterogeneous constraining power of individual systems, where some mergers contribute substantially more information than others. Notably, the shifts produced by all four prior variations remain well within this bootstrap range. We therefore conclude that while prior dependence is observable, it is subdominant to the sample-to-sample variability that governs the current error budget.}

\subsection{Per-System Contributions to the Joint Constraint}\label{sec:per_system}

\edit1{The heterogeneous constraining power noted above raises the question of whether the joint constraint in Figure~\ref{fig:sigma_m_real} is dominated by a small number of systems. To address this, we computed individual upper limits from each system's marginalized posterior (the faint lines in Figure~\ref{fig:sigma_m_real}). The individual constraints are uniformly weak: the per-system 68\% (95\%) upper limits range from 0.7 to 1.5 (1.6 to 1.9)~$\text{cm}^{2}\,\text{g}^{-1}$ (Table~\ref{tab:per_system}), with the 95\% values only modestly below the prior boundary of $2~\text{cm}^{2}\,\text{g}^{-1}$. Even the most constraining single system is thus a factor of $\sim$3.2 weaker than the joint 68\% limit of $0.22~\text{cm}^{2}\,\text{g}^{-1}$ ($0.63~\text{cm}^{2}\,\text{g}^{-1}$ at 95\%), demonstrating that the result is a genuine ensemble property rather than the imprint of any individual cluster, consistent with the $1/\sqrt{N}$ scaling seen in our mock recovery tests (Figure~\ref{fig:sigma_m_multi}).}

\begin{deluxetable*}{lccc}
\tablecaption{\editcap{Per-System Time-Since-Collision and Jackknife $\sigma/m$ Upper Limits \label{tab:per_system}}}
\tablecolumns{4}
\tablewidth{0pt}
\tablehead{
\colhead{Name} & \colhead{$t_{sc}$ [Gyr]} & \colhead{Individual $\sigma/m$} & \colhead{Jackknife $\sigma/m$} \\
\colhead{} & \colhead{} & \colhead{68\% (95\%)} & \colhead{68\% (95\%)}
}
\startdata
MACS J1752.0+4440  & $0.44^{+0.03}_{-0.02}$ & 0.70 (1.64) & 0.38 (0.95) \\
Abell 3376         & $0.58^{+0.06}_{-0.06}$ & 0.79 (1.70) & 0.33 (0.92) \\
Abell 1240         & $0.73^{+0.05}_{-0.05}$ & 0.95 (1.75) & 0.29 (0.82) \\
ZWCL 0008.8+5215   & $0.45^{+0.03}_{-0.03}$ & 1.03 (1.81) & 0.29 (0.80) \\
Abell 2345         & $0.58^{+0.04}_{-0.04}$ & 1.07 (1.82) & 0.27 (0.77) \\
ZWCL 1856.8+6616   & $0.66^{+0.07}_{-0.06}$ & 1.19 (1.86) & 0.25 (0.68) \\
PSZ2 G181.06+48.47 & $0.95^{+0.06}_{-0.06}$ & 1.34 (1.90) & 0.23 (0.63) \\
CIZA J2242.8+5301  & $0.50^{+0.04}_{-0.04}$ & 1.38 (1.90) & 0.24 (0.64) \\
ACT-CL J0102-4915  & $0.42^{+0.02}_{-0.01}$ & 1.41 (1.90) & 0.22 (0.62) \\
RXC J1314.4-2515   & $0.45^{+0.04}_{-0.03}$ & 1.47 (1.92) & 0.22 (0.59) \\
ZWCL1447.2+2619    & $0.73^{+0.07}_{-0.06}$ & 1.51 (1.93) & 0.21 (0.59) \\
\hline
Joint (all 11)     & \nodata & 0.22 (0.63) & \nodata \\
\enddata
\tablecomments{\editcap{Upper limits on $\sigma/m$ are in units of $\text{cm}^{2}\,\text{g}^{-1}$. The $t_{sc}$ column gives the time since collision (median with $1\sigma$ uncertainties, in Gyr) pinned down by the shock-to-shock distance acting as a merger chronometer. The ``Individual'' column gives the 68\% (95\%) upper limit derived from each system's marginalized posterior; the ``Jackknife'' column gives the joint 68\% (95\%) limit obtained when that system is removed from the gold sample. Systems are ordered from most to least constraining. The bottom row lists the joint limit from all eleven systems.}}
\end{deluxetable*}

\edit1{The systems nevertheless contribute unequally. The most informative is MACS~J1752.0+4440, with an individual 68\% upper limit of $0.70~\text{cm}^{2}\,\text{g}^{-1}$, followed by Abell~3376 ($0.79~\text{cm}^{2}\,\text{g}^{-1}$) and Abell~1240 ($0.95~\text{cm}^{2}\,\text{g}^{-1}$). We suspect that this ranking has a physical origin: MACS~J1752.0+4440 combines the largest observed halo-to-halo separation with the smallest fractional uncertainty ($1.19 \pm 0.07$~Mpc; Table~\ref{tab:gold}), and a large halo separation at a given shock separation most strongly disfavors the SIDM-induced drag. A jackknife test (Table~\ref{tab:per_system}) confirms that the joint result does not hinge on this single system: removing MACS~J1752.0+4440 relaxes the joint 68\% limit from 0.22 to $0.38~\text{cm}^{2}\,\text{g}^{-1}$ (0.63 to 0.95 at 95\%), the largest change produced by excluding any single system, while still remaining below the $\sigma/m \gtrsim 1~\text{cm}^{2}\,\text{g}^{-1}$ regime invoked to address small-scale structure problems. The jackknife is also two-sided: excluding the least constraining systems (e.g., ZWCL~1447.2+2619, whose individual posterior mildly rises toward larger $\sigma/m$, perhaps owing to its small observed halo separation) slightly tightens the joint limit, indicating that the ensemble contains systems pulling in both directions rather than being uniformly stacked against large cross-sections.}

\begin{figure*}
\centering
\includegraphics[width=0.85\textwidth]{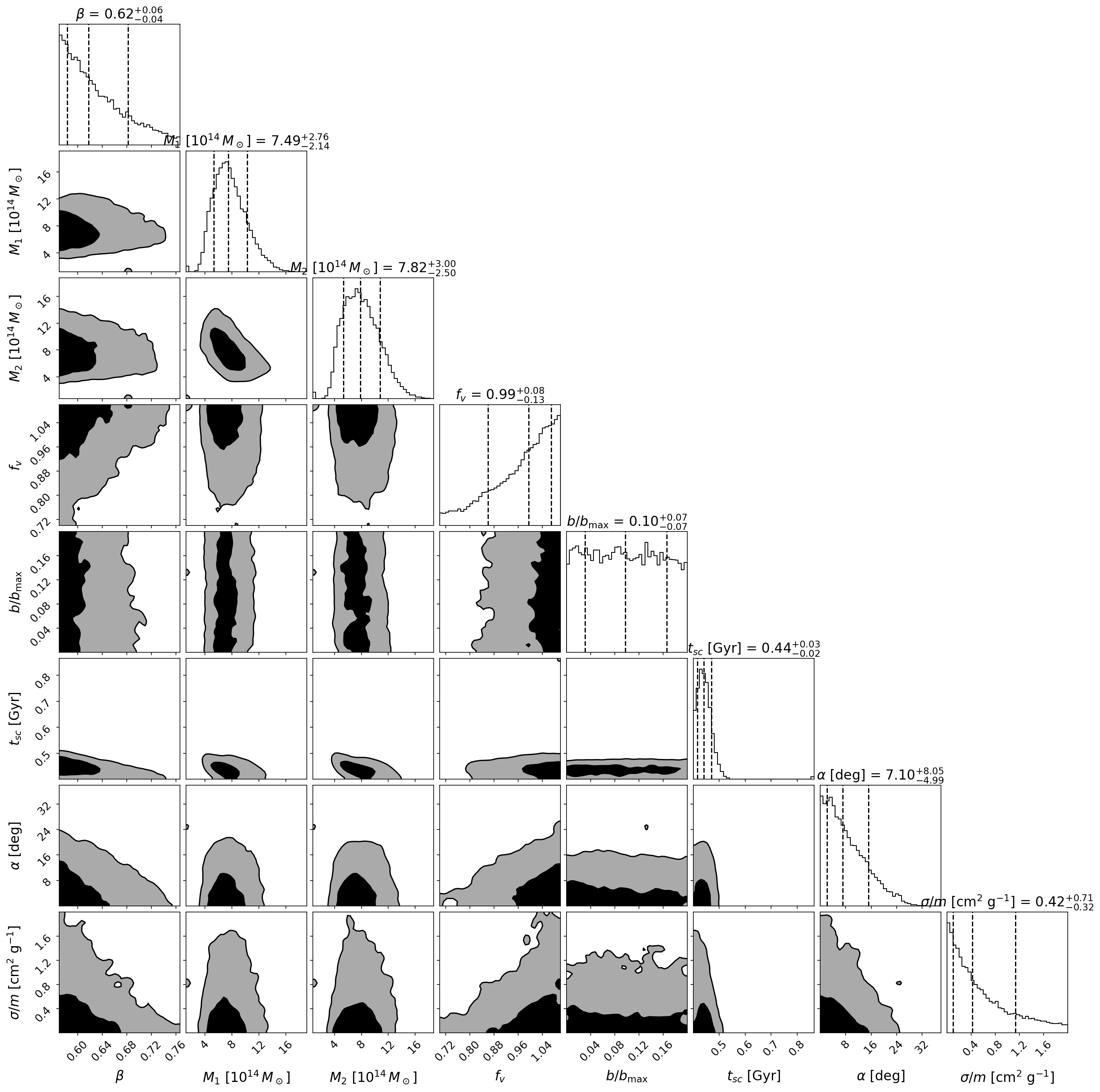}
\caption{\editcap{Posterior distribution of all eight model parameters for MACS~J1752.0+4440, the most constraining system in our gold sample. The contours indicate the 1$\sigma$ and 2$\sigma$ credible regions, and the dashed lines in the marginal histograms mark the 16th, 50th, and 84th percentiles. No single nuisance parameter is strongly degenerate with $\sigma/m$; rather, mild correlations with several parameters collectively broaden the $\sigma/m$ posterior, illustrating why marginalization over the full parameter space and the combination of independent systems are essential for a robust SICS constraint.}}
\label{fig:corner_macs1752}
\end{figure*}

\edit1{Figure~\ref{fig:corner_macs1752} presents the full eight-dimensional posterior for MACS~J1752.0+4440, showing the marginalized distributions and mutual correlations of all sampled parameters. Because the model contains eight free parameters while each system supplies only two scalar constraints (the projected shock-to-shock and halo-to-halo distances), most marginalized posteriors are necessarily prior-dominated and not very informative. The notable exception is the merger phase: the shock-to-shock distance acts as a merger chronometer and pins down $t_{sc} = 0.44^{+0.03}_{-0.02}$~Gyr, a factor of $\sim$6 tighter than the prior. No single nuisance parameter exhibits a strong degeneracy with $\sigma/m$; instead, mild correlations, most notably with the halo masses and the merger phase, collectively broaden the $\sigma/m$ posterior. This illustrates why an individual system, however well measured, cannot tightly constrain the SICS on its own, and why marginalization over the full nuisance space, followed by the combination of independent systems, is essential to our approach.}

\subsection{Impact of the Mass--Concentration Relation}\label{sec:concentration}

\edit1{Our simulation grid assigns halo concentrations through the mass--concentration ($M$--$c$) relation of \citet{Duffy2008}, and one may ask whether departures from this relation, either through the intrinsic scatter in concentration or through changes in the functional form of the density profile, could mimic the SIDM-induced suppression of the halo separation (Figure~\ref{fig:t_vs_d}) and thus be degenerate with $\sigma/m$. Incorporating a full marginalization over the $M$--$c$ scatter into the simulation grid is computationally prohibitive: the current set of 576 runs requires approximately three months on our computing facility, and adding even two runs per merger to bracket the $\pm1\sigma$ concentration boundaries would roughly triple the total. We therefore performed dedicated perturbation tests to directly estimate the associated systematic.}

\edit1{Since the scatter in the $M$--$c$ relation is approximately log-normal, we adopt $\delta \log_{10} c = 0.11$ and generate randomized initial conditions for the eleven gold systems in Table~\ref{tab:gold}, assuming the \citet{Duffy2008} $M$--$c$ relation. When the concentrations of both halos are increased by $1\sigma$, the shock propagation speed increases by $\sim$1\%, while the maximum halo-to-halo separation decreases by 2--4\%, as expected from the enhanced dynamical friction experienced by more concentrated halos. These offsets are within the observational measurement uncertainties (Table~\ref{tab:gold}) and are thus subdominant in the $\sigma/m$ constraint: the resulting shift of the 68\% (95\%) upper limit is $\sim0.05~\text{cm}^{2}\,\text{g}^{-1}$ ($\sim0.03~\text{cm}^{2}\,\text{g}^{-1}$). }

\edit1{More generally, any departure of our idealized setup from reality, including deviations in the halo profile shape, triaxiality, substructure, mass accretion, and surrounding large-scale structure, can in principle modify the trend shown in Figure~\ref{fig:t_vs_d}. While expanding the simulation grid to marginalize over all of these effects is beyond the scope of the current study, the TNG-Cluster validation (\S\ref{sec:tngcluster}) plays a critical role in this regard: it demonstrates that our method recovers consistent posteriors when applied to clusters drawn from a cosmological zoom-in simulation, in which the full diversity of halo profiles and environments is naturally present.}

\subsection{Future Outlook: Synergies with Rubin, Roman, Euclid, and SKA}\label{sec:future}
The methodology established in this study provides a scalable framework for the next decade of cosmological surveys. As the census of known radio relic clusters is expected to grow from $\sim$120 to a few thousand with the full operation of the Square Kilometre Array (SKA) \citep{Lee2026}, the statistical uncertainty on $\sigma/m$ is projected to decrease as $1/\sqrt{N}$ (Figure~\ref{fig:sigma_m_multi}). However, this transition to a population-wide statistical analysis will require high-precision imaging and multi-wavelength capabilities from upcoming flagship missions to control systematic floors.

Future extensions of this work will leverage the following advancements:
\begin{itemize}
\item \textbf{High-Resolution Weak Lensing with Roman and Euclid:} The sub-arcsecond resolution of the Nancy Grace Roman Space Telescope and Euclid will provide a stable, diffraction-limited Point Spread Function (PSF). This will drastically reduce measurement uncertainties for halo-to-halo distances ($d_{h}$) by resolving smaller, high-redshift background galaxies and controlling shear calibration biases.
\item \textbf{Advanced Cosmological Modeling with TNGC-SIDM:} While the current study relies on idealized simulations, future work will utilize the TNG-Cluster-SIDM (TNGC-SIDM) simulations (W. Lee et al., in prep), which will implement SIDM models within the merging cluster systems identified in TNG-Cluster. These specialized cosmological zoom-in simulations will track the evolution of multi-wavelength observables in merging clusters with dark matter self-interaction, providing a direct testbed for our proposed method. As a preliminary validation, a low-mass cluster merger analog confirms that the shock separation evolves independently of the SICS (W. Lee et al., in prep), supporting the robustness of our indicator even under real-life complexities such as large-scale structures, mass accretion, and halo substructures.
\item \textbf{Line-of-Sight Kinematics:} Utilizing next-generation spectroscopic surveys will allow for the integration of line-of-sight velocity information for cluster member galaxies. This added dimension of phase-space data can break the current degeneracy between the viewing angle ($\alpha$) and the collision velocity ($f_{v}$), both of which are currently treated as marginalized nuisance parameters in our MCMC framework. Beyond optical spectroscopy, the kinematic Sunyaev--Zel'dovich (kSZ) effect provides an independent line-of-sight velocity probe of the bulk gas motion in merging clusters. Although no double-relic system has yet been mapped this way, pilot studies such as the ICM-SHOX program have already demonstrated kSZ-based gas--dark matter velocity decoupling in a cluster merger \citep{Silich2024}, and forthcoming wide-field submillimetre facilities such as AtLAST are expected to deliver routine cluster kSZ measurements.
\item \textbf{Independent Viewing-Angle Constraints from Radio Polarization:} The polarization fraction and orientation of double radio relics encode the projected geometry of the shock fronts and can independently constrain the merger viewing angle $\alpha$. Combining polarimetric data from upcoming SKA-era surveys with the kinematic framework presented here will further reduce the projection-related systematic floor.
\item \textbf{Enhanced Radio Selection via SKA:} The increased sensitivity of the SKA will allow for the detection of more symmetric, gold-standard double radio relics. By aligning these radio-traced shock positions with high-precision mass maps from Roman and Euclid, we can further refine the use of the relic-relic distance as a robust merger chronometer for dark matter physics. Moreover, SKA's sensitivity will preferentially expand the census toward fainter, late-phase relics, where the cumulative effect of self-interaction on halo deceleration is largest and the SIDM/CDM contrast in the $d_h/d_s$ ratio is most pronounced. This phase-space extension will yield not only more systems but also higher per-system constraining power.
\end{itemize}
By combining the statistical power of the Rubin Observatory and SKA with the high-resolution imaging of Roman and Euclid, supported by the cosmological realism of TNGC-SIDM, future studies will be able to probe the velocity-dependence of the dark matter cross-section with unprecedented precision.

\section{Conclusion}\label{sec:conclusion}
In this paper, we have presented a new and robust constraint on the dark matter self-interaction cross-section ($\sigma/m$) using a sample of eleven ``gold'' cluster collisions hosting double radio relics. By utilizing the shock-to-shock distance as a dynamical anchor, we have overcome the systematic uncertainties and geometric ambiguities that have historically hindered measurements based on mass-galaxy offsets. Our joint-likelihood analysis, marginalized over mass, viewing angle, collision velocity, merger phase, and impact parameter, yields \edit1{an upper limit of $\sigma/m < 0.22~(0.63)~\text{cm}^{2}\,\text{g}^{-1}$ at the 68\% (95\%) confidence level}. This represents the most stringent constraint obtained to date from cluster-scale collisions.

These results place our constraint in mild tension with existing SIDM models that require $\sigma/m \gtrsim 1~\text{cm}^{2}\,\text{g}^{-1}$ to resolve small-scale discrepancies in dwarf and spiral galaxies \citep{Zavala2013, Elbert2015, Kaplinghat2016}. This discrepancy suggests that if dark matter self-interaction is indeed the solution to these puzzles, its cross-section must be significantly suppressed at the high-velocity scales typical of cluster mergers, a behavior naturally predicted by velocity-dependent SIDM theories \citep{Tulin2013, Ackerman2009, Boddy2014}. While the current sample size is insufficient to explicitly map this velocity dependence, our results provide a robust ``effective'' upper limit within the rare-scattering (contact-interaction) regime. From our mock recovery tests (Figure~\ref{fig:sigma_m_multi}), a sample of $\sim 100$ double radio relic systems would tighten this limit to the $\sim 0.1~\text{cm}^{2}\,\text{g}^{-1}$ level, at which velocity-dependent models become meaningfully distinguishable. The applicability of this framework to more general self-interaction models, including the long-range frequent-interactions, will be explored in future work.

The robustness of our methodology is supported by extensive validation against both idealized and cosmological (TNG-Cluster) simulations. These tests confirm that stacking independent merger systems effectively averages out system-specific nuisance parameters, such as merger phase and viewing angle, allowing for a precise recovery of the true cross-section despite the inherent ambiguities of individual clusters.

As the census of radio relic clusters grows with next-generation surveys like the SKA, this framework will enable increasingly precise probes of dark-sector physics. By combining the statistical power of upcoming radio and optical surveys, we will soon be able to directly map the velocity dependence of dark matter self-interaction in regimes inaccessible to terrestrial laboratories.

\begin{acknowledgments}
We thank the Merging Cluster Collaboration for their useful comments. MJJ acknowledges support for the current research from the National Research Foundation (NRF) of Korea under the programs 2022R1A2C1003130 and RS-2023-00219959.
\end{acknowledgments}

\bibliographystyle{aasjournalv7}
\bibliography{ms}

\begin{thebibliography}{}
\expandafter\ifx\csname natexlab\endcsname\relax\def\natexlab#1{#1}\fi
\providecommand{\url}[1]{\href{#1}{#1}}
\providecommand{\dodoi}[1]{doi:~\href{http://doi.org/#1}{\nolinkurl{#1}}}
\providecommand{\doeprint}[1]{\href{http://ascl.net/#1}{\nolinkurl{http://ascl.net/#1}}}
\providecommand{\doarXiv}[1]{\href{https://arxiv.org/abs/#1}{\nolinkurl{https://arxiv.org/abs/#1}}}

\bibitem[{L. {Ackerman} {et~al.}(2009){Ackerman}, {Buckley}, {Carroll}, \&
  {Kamionkowski}}]{Ackerman2009}
{Ackerman}, L., {Buckley}, M.~R., {Carroll}, S.~M., \& {Kamionkowski}, M. 2009,
  \bibinfo{title}{{Dark matter and dark radiation},} Phys. Rev. D, 79, 023519,
  \dodoi{10.1103/PhysRevD.79.023519}

\bibitem[{E. Ahn {et~al.}(2025)Ahn {et~al.}}]{Ahn2025}
Ahn, E., {et~al.} 2025, \bibinfo{title}{{A Multiwavelength Study of the
  High-redshift Double Radio Relic Cluster PSZ2 G181.06+48.47},} Astrophys. J.,
  984, 26, \dodoi{10.3847/1538-4357/ad0f1e}

\bibitem[{A.~J. {Benson}(2005){Benson}}]{Benson2005}
{Benson}, A.~J. 2005, \bibinfo{title}{{Orbital parameters of infalling dark
  matter substructures},} \mnras, 358, 551,
  \dodoi{10.1111/j.1365-2966.2005.08788.x}

\bibitem[{K.~K. {Boddy} {et~al.}(2014){Boddy}, {Feng}, {Kaplinghat}, \&
  {Tait}}]{Boddy2014}
{Boddy}, K.~K., {Feng}, J.~L., {Kaplinghat}, M., \& {Tait}, T.~M.~P. 2014,
  \bibinfo{title}{{Self-interacting dark matter from a non-Abelian hidden
  sector},} Phys. Rev. D, 89, 115017, \dodoi{10.1103/PhysRevD.89.115017}

\bibitem[{H. Cho {et~al.}(2022)Cho {et~al.}}]{Cho2022}
Cho, H., {et~al.} 2022, \bibinfo{title}{{Multiwavelength analysis of A1240, the
  double radio-relic merging galaxy cluster embedded in an 80 Mpc-long cosmic
  filament},} Astrophys. J., 925, 68, \dodoi{10.3847/1538-4357/ac3a00}

\bibitem[{A.~R. {Duffy} {et~al.}(2008){Duffy}, {Schaye}, {Kay}, \& {Dalla
  Vecchia}}]{Duffy2008}
{Duffy}, A.~R., {Schaye}, J., {Kay}, S.~T., \& {Dalla Vecchia}, C. 2008,
  \bibinfo{title}{{Dark matter halo concentrations in the Wilkinson Microwave
  Anisotropy Probe year 5 cosmology},} \mnras, 390, L64,
  \dodoi{10.1111/j.1745-3933.2008.00537.x}

\bibitem[{O.~D. {Elbert} {et~al.}(2015){Elbert} {et~al.}}]{Elbert2015}
{Elbert}, O.~D., {et~al.} 2015, \bibinfo{title}{{Core formation in dwarf haloes
  with self-interacting dark matter: no fine-tuning necessary},} Mon. Not. R.
  Astron. Soc., 453, 29, \dodoi{10.1093/mnras/stv1470}

\bibitem[{T.~A. {Ensslin} {et~al.}(1998){Ensslin}, {Biermann}, {Klein}, \&
  {Kohle}}]{Ensslin1998}
{Ensslin}, T.~A., {Biermann}, P.~L., {Klein}, U., \& {Kohle}, S. 1998,
  \bibinfo{title}{{Cluster radio relics as a tracer of shock waves of the
  large-scale structure formation},} \aap, 332, 395

\bibitem[{L. {Feretti} {et~al.}(2012){Feretti}, {Giovannini}, {Govoni}, \&
  {Murgia}}]{Feretti2012}
{Feretti}, L., {Giovannini}, G., {Govoni}, F., \& {Murgia}, M. 2012,
  \bibinfo{title}{{Clusters of galaxies: observational properties of the
  diffuse radio emission},} The Astronomy and Astrophysics Review, 20, 54,
  \dodoi{10.1007/s00159-012-0054-z}

\bibitem[{K. {Finner} {et~al.}(2025){Finner}, {Jee}, {Cho}, {HyeongHan}, {Lee},
  {van Weeren}, {Wittman}, \& {Yoon}}]{Finner2025}
{Finner}, K., {Jee}, M.~J., {Cho}, H., {et~al.} 2025,
  \bibinfo{title}{{Weak-lensing Characterization of the Dark Matter in 29
  Merging Clusters that Exhibit Radio Relics},} \apjs, 277, 28,
  \dodoi{10.3847/1538-4365/adb0b6}

\bibitem[{M.~S. {Fischer} {et~al.}(2023){Fischer}, {Durke}, {Hollingshausen},
  {Hammer}, {Br{\"u}ggen}, \& {Dolag}}]{Fischer2023}
{Fischer}, M.~S., {Durke}, N.-H., {Hollingshausen}, K., {et~al.} 2023,
  \bibinfo{title}{{The role of baryons in self-interacting dark matter
  mergers},} \mnras, 523, 5915, \dodoi{10.1093/mnras/stad1786}

\bibitem[{D. {Foreman-Mackey} {et~al.}(2013){Foreman-Mackey}, {Hogg}, {Lang},
  \& {Goodman}}]{ForemanMackey2013}
{Foreman-Mackey}, D., {Hogg}, D.~W., {Lang}, D., \& {Goodman}, J. 2013,
  \bibinfo{title}{{emcee: The MCMC Hammer},} \pasp, 125, 306,
  \dodoi{10.1086/670067}

\bibitem[{N. {Golovich} {et~al.}(2019){Golovich}, {Dawson}, {Wittman},
  {Ogrean}, {van Weeren}, \& {Bonafede}}]{Golovich2019}
{Golovich}, N., {Dawson}, W.~A., {Wittman}, D., {et~al.} 2019,
  \bibinfo{title}{{Merging Cluster Collaboration: Optical and Spectroscopic
  Survey of a Radio-selected Sample of 29 Merging Galaxy Clusters},} \apjs,
  240, 39, \dodoi{10.3847/1538-4365/aaf88b}

\bibitem[{J.-H. {Ha} {et~al.}(2018){Ha}, {Ryu}, \& {Kang}}]{Ha2018}
{Ha}, J.-H., {Ryu}, D., \& {Kang}, H. 2018, \bibinfo{title}{{Properties of
  Merger Shocks in Merging Galaxy Clusters},} \apj, 857, 26,
  \dodoi{10.3847/1538-4357/aab3e6}

\bibitem[{D. Harvey {et~al.}(2015)Harvey, Massey, Kitching, Taylor, \&
  Tittley}]{Harvey2015}
Harvey, D., Massey, R., Kitching, T., Taylor, A., \& Tittley, E. 2015,
  \bibinfo{title}{The nongravitational interactions of dark matter in colliding
  galaxy clusters,} Science, 347, 1462

\bibitem[{P.~F. Hopkins(2015)Hopkins}]{Hopkins2015}
Hopkins, P.~F. 2015, \bibinfo{title}{A new class of accurate, mesh-free
  hydrodynamic simulation methods,} Mon. Not. R. Astron. Soc., 450, 53,
  \dodoi{10.1093/mnras/stv195}

\bibitem[{M.~J. Jee {et~al.}(2015)Jee {et~al.}}]{Jee2015}
Jee, M.~J., {et~al.} 2015, \bibinfo{title}{{$MC^{2}$: Constraining the dark
  matter distribution of the violent merging galaxy cluster CIZA J2242.8+5301
  by piercing through the Milky Way},} Astrophys. J., 802, 46,
  \dodoi{10.1088/0004-637X/802/1/46}

\bibitem[{C. {Jones} \& W. {Forman}(1984){Jones} \& {Forman}}]{Jones1984}
{Jones}, C., \& {Forman}, W. 1984, \bibinfo{title}{{The structure of clusters
  of galaxies observed with Einstein},} \apj, 276, 38, \dodoi{10.1086/161591}

\bibitem[{F. Kahlhoefer {et~al.}(2014)Kahlhoefer, Schmidt-Hoberg, Frandsen, \&
  Sarkar}]{Kahlhoefer2014}
Kahlhoefer, F., Schmidt-Hoberg, K., Frandsen, M., \& Sarkar, S. 2014,
  \bibinfo{title}{Colliding clusters and dark matter self-interactions,} Mon.
  Not. R. Astron. Soc., 437, 2865, \dodoi{10.1093/mnras/stt2097}

\bibitem[{H. {Kang} {et~al.}(2012){Kang}, {Ryu}, \& {Jones}}]{Kang2012}
{Kang}, H., {Ryu}, D., \& {Jones}, T.~W. 2012, \bibinfo{title}{{Diffusive Shock
  Acceleration Simulations of Radio Relics in Galaxy Clusters},} \apj, 756, 97,
  \dodoi{10.1088/0004-637X/756/1/97}

\bibitem[{M. {Kaplinghat} {et~al.}(2016){Kaplinghat}, {Tulin}, \&
  {Yu}}]{Kaplinghat2016}
{Kaplinghat}, M., {Tulin}, S., \& {Yu}, H. 2016, \bibinfo{title}{{Dark Matter
  Halos as Particle Colliders: Unified Solution to Small-Scale Structure
  Puzzles from Dwarfs to Clusters},} Phys. Rev. Lett., 116, 041302,
  \dodoi{10.1103/PhysRevLett.116.041302}

\bibitem[{J. Kim {et~al.}(2021)Kim {et~al.}}]{Kim2021}
Kim, J., {et~al.} 2021, \bibinfo{title}{{Head-to-toe measurement of El Gordo:
  improved analysis of the galaxy cluster ACT-CL J0102-4915 with new wide-field
  Hubble space telescope imaging data},} Astrophys. J., 923, 101,
  \dodoi{10.3847/1538-4357/ac294d}

\bibitem[{S.~Y. Kim {et~al.}(2017)Kim, Peter, \& Wittman}]{Kim2017}
Kim, S.~Y., Peter, A. H.~G., \& Wittman, D. 2017, \bibinfo{title}{In the wake
  of dark giants: new signatures of dark matter self-interactions in equal-mass
  mergers of galaxy clusters,} Mon. Not. R. Astron. Soc., 469, 1414,
  \dodoi{10.1093/mnras/stx896}

\bibitem[{K. {Knowles} {et~al.}(2022){Knowles}, {Cotton}, {Rudnick}, {Camilo},
  {Gozaliasl}, {Grover}, {Heald}, {Heywood}, {Huffenberger}, {Jarvis}, {Jones},
  {Khatry}, {Ku{\v{z}}mi{\'c}}, {Loubser}, {Mauch}, {Murphy}, {Mroczkowski},
  {Parekh}, {Perley}, {Ramatsoku}, {Sammons}, {Sikhosana}, {Smirnov}, {Thorat},
  \& {van derHeyden}}]{Knowles2022}
{Knowles}, K., {Cotton}, W.~D., {Rudnick}, L., {et~al.} 2022,
  \bibinfo{title}{{The MeerKAT Galaxy Cluster Legacy Survey. I. Survey Overview
  and Highlights},} Astronomy \& Astrophysics, 657, A56,
  \dodoi{10.1051/0004-6361/202141488}

\bibitem[{W. {Lee} {et~al.}(2026){Lee}, {Pillepich}, {Nelson}, {Jee}, {Nagai},
  {Finner}, \& {ZuHone}}]{Lee2026}
{Lee}, W., {Pillepich}, A., {Nelson}, D., {et~al.} 2026,
  \bibinfo{title}{{Exploring the Statistical Properties of Double Radio Relics
  in the TNG-Cluster and TNG300 Simulations},} \apj, 998, 201,
  \dodoi{10.3847/1538-4357/ae3245}

\bibitem[{W. {Lee} {et~al.}(2024){Lee}, {Pillepich}, {ZuHone}, {Nelson}, {Jee},
  {Nagai}, \& {Finner}}]{Lee2024}
{Lee}, W., {Pillepich}, A., {ZuHone}, J., {et~al.} 2024, \bibinfo{title}{{Radio
  relics in massive galaxy cluster mergers in the TNG-Cluster simulation},}
  \aap, 686, A55, \dodoi{10.1051/0004-6361/202348194}

\bibitem[{W. Lee {et~al.}(2022)Lee {et~al.}}]{Lee2022}
Lee, W., {et~al.} 2022, \bibinfo{title}{{Discovery of a double radio relic in
  ZWCL1447.2+2619: a rare testbed for shock-acceleration models with a peculiar
  surface-brightness ratio},} Astrophys. J., 924, 18,
  \dodoi{10.3847/1538-4357/ac35e4}

\bibitem[{M. Markevitch {et~al.}(2004)Markevitch {et~al.}}]{Markevitch2004}
Markevitch, M., {et~al.} 2004, \bibinfo{title}{Direct Constraints on the Dark
  Matter Self-Interaction Cross Section from the Merging Galaxy Cluster 1E
  0657-56,} Astrophys. J., 606, 819

\bibitem[{J.~J. {Mohr} {et~al.}(1999){Mohr}, {Mathiesen}, \&
  {Evrard}}]{Mohr1999}
{Mohr}, J.~J., {Mathiesen}, B., \& {Evrard}, A.~E. 1999,
  \bibinfo{title}{{Properties of the Intracluster Medium in an Ensemble of
  Nearby Clusters},} \apj, 517, 627, \dodoi{10.1086/307197}

\bibitem[{R. {Monteiro-Oliveira} {et~al.}(2017){Monteiro-Oliveira}
  {et~al.}}]{Monteiro-Oliveira2017}
{Monteiro-Oliveira}, R., {et~al.} 2017, \bibinfo{title}{{Weak lensing and
  spectroscopic analysis of the nearby dissociative merging galaxy cluster
  Abell 3376},} Mon. Not. R. Astron. Soc., 468, 4566,
  \dodoi{10.1093/mnras/stx791}

\bibitem[{J.~F. Navarro {et~al.}(1997)Navarro, Frenk, \& White}]{NFW1997}
Navarro, J.~F., Frenk, C.~S., \& White, S. D.~M. 1997, \bibinfo{title}{A
  Universal Density Profile from Hierarchical Clustering,} Astrophys. J., 490,
  493, \dodoi{10.1086/304888}

\bibitem[{D. {Nelson} {et~al.}(2024){Nelson}, {Pillepich}, {Ayromlou}, {Lee},
  {Lehle}, {Rohr}, \& {Truong}}]{Nelson2024}
{Nelson}, D., {Pillepich}, A., {Ayromlou}, M., {et~al.} 2024,
  \bibinfo{title}{{Introducing the TNG-Cluster simulation: Overview and the
  physical properties of the gaseous intracluster medium},} \aap, 686, A157,
  \dodoi{10.1051/0004-6361/202348608}

\bibitem[{D. {Park} \& M.~J. {Jee}(2022){Park} \& {Jee}}]{Park2022}
{Park}, D., \& {Jee}, M.~J. 2022, \bibinfo{title}{{Toward Robust Reconstruction
  of Cluster Merger Scenarios},} BAAS, 54, 139.17

\bibitem[{C. Power {et~al.}(2003)Power, Navarro, Jenkins, Frenk, White,
  Springel, Stadel, \& Quinn}]{Power2003}
Power, C., Navarro, J.~F., Jenkins, A., {et~al.} 2003, \bibinfo{title}{The
  inner structure of {$\Lambda$}CDM haloes - I. A numerical convergence study,}
  Mon. Not. R. Astron. Soc., 338, 14, \dodoi{10.1046/j.1365-8711.2003.05925.x}

\bibitem[{K. {Rajpurohit} {et~al.}(2025){Rajpurohit}, {Stroe}, {O'Sullivan},
  {Ahn}, {Lee}, {Cho}, {Jee}, {van Weeren}, {et~al.}}]{Rajpurohit2025}
{Rajpurohit}, K., {Stroe}, A., {O'Sullivan}, E., {et~al.} 2025,
  \bibinfo{title}{{PSZ2 G181.06+48.47 II: radio analysis of a low-mass cluster
  with exceptionally-distant radio relics},} Astrophys. J., 984, 25,
  \dodoi{10.3847/1538-4357/adbbb9}

\bibitem[{M. {Rocha} {et~al.}(2013){Rocha}, {Peter}, {Bullock}, {Kaplinghat},
  {Garrison-Kimmel}, {O{\~n}orbe}, \& {Moustakas}}]{Rocha2013}
{Rocha}, M., {Peter}, A.~H.~G., {Bullock}, J.~S., {et~al.} 2013,
  \bibinfo{title}{{Cosmological simulations with self-interacting dark matter
  -- I. Constant-density cores and substructure},} \mnras, 430, 81,
  \dodoi{10.1093/mnras/sts514}

\bibitem[{C.~L. Sarazin(2002)Sarazin}]{Sarazin2002}
Sarazin, C.~L. 2002, \bibinfo{title}{The Physics of Cluster Mergers,} in
  Astrophysics and Space Science Library, Vol. 272, Merging Processes in
  Clusters of Galaxies, ed. L.~Feretti, I.~M. Gioia, \& G.~Giovannini
  (Dordrecht: Kluwer Academic Publishers), 1--38,
  \dodoi{10.1007/0-306-48096-4_1}

\bibitem[{E.~M. {Silich} {et~al.}(2024){Silich}, {Bellomi}, {Sayers}, {ZuHone},
  {Chadayammuri}, {Golwala}, {Hughes}, {Monta{\~n}a}, {Mroczkowski}, {Nagai},
  {S{\'a}nchez-Arg{\"u}elles}, {Stanford}, {Wilson}, {Zemcov}, \&
  {Zitrin}}]{Silich2024}
{Silich}, E.~M., {Bellomi}, E., {Sayers}, J., {et~al.} 2024,
  \bibinfo{title}{{ICM-SHOX. I. Methodology Overview and Discovery of a
  Gas-Dark Matter Velocity Decoupling in the MACS J0018.5+1626 Merger},} \apj,
  968, 74, \dodoi{10.3847/1538-4357/ad3fb5}

\bibitem[{D.~N. Spergel \& P.~J. Steinhardt(2000)Spergel \&
  Steinhardt}]{Spergel2000}
Spergel, D.~N., \& Steinhardt, P.~J. 2000, \bibinfo{title}{Observational
  evidence for self-interacting cold dark matter,} Phys. Rev. Lett., 84, 3760

\bibitem[{S. Tulin \& H. Yu(2018)Tulin \& Yu}]{Tulin2018}
Tulin, S., \& Yu, H. 2018, \bibinfo{title}{Dark matter self-interactions and
  small scale structure,} Phys. Rep., 730, 1

\bibitem[{S. Tulin {et~al.}(2013)Tulin, Yu, \& Zurek}]{Tulin2013}
Tulin, S., Yu, H.-B., \& Zurek, K.~M. 2013, \bibinfo{title}{Beyond
  collisionless dark matter: Particle physics dynamics for dark matter halo
  structure,} Phys. Rev. D, 87, 115007, \dodoi{10.1103/PhysRevD.87.115007}

\bibitem[{R.~J. {van Weeren} {et~al.}(2019){van Weeren}, {de Gasperin},
  {Akamatsu}, {Br{\"u}ggen}, {Feretti}, {Kang}, {Stroe}, \&
  {Zandanel}}]{vanWeeren2019}
{van Weeren}, R.~J., {de Gasperin}, F., {Akamatsu}, H., {et~al.} 2019,
  \bibinfo{title}{{Diffuse Radio Emission from Galaxy Clusters},} Space Science
  Reviews, 215, 16, \dodoi{10.1007/s11214-019-0584-z}

\bibitem[{R.~J. {van Weeren} {et~al.}(2012){van Weeren}, {R{\"o}ttgering},
  {Intema}, {Rudnick}, {Br{\"u}ggen}, {Hoeft}, \& {Oonk}}]{vanWeeren2011}
{van Weeren}, R.~J., {R{\"o}ttgering}, H.~J.~A., {Intema}, H.~T., {et~al.}
  2012, \bibinfo{title}{{The ``Toothbrush-relic'': evidence for a coherent
  linear 2-Mpc scale shock wave in a massive merging galaxy cluster?},} \aap,
  546, A124, \dodoi{10.1051/0004-6361/201219000}

\bibitem[{A.~R. {Wetzel}(2011){Wetzel}}]{Wetzel2011}
{Wetzel}, A.~R. 2011, \bibinfo{title}{{On the orbits of infalling satellite
  haloes},} \mnras, 412, 49, \dodoi{10.1111/j.1365-2966.2010.17877.x}

\bibitem[{D. Wittman {et~al.}(2018)Wittman, Golovich, \& Dawson}]{Wittman2018}
Wittman, D., Golovich, N., \& Dawson, W.~A. 2018, \bibinfo{title}{The
  mismeasure of mergers: revised limits on self-interacting dark matter in
  merging galaxy clusters,} Astrophys. J., 869, 104

\bibitem[{J. {Zavala} {et~al.}(2013){Zavala}, {Vogelsberger}, \&
  {Walker}}]{Zavala2013}
{Zavala}, J., {Vogelsberger}, M., \& {Walker}, M.~G. 2013,
  \bibinfo{title}{{Constraining self-interacting dark matter with the Milky
  Way's dwarf spheroidals},} Mon. Not. R. Astron. Soc. Lett., 431, L20,
  \dodoi{10.1093/mnrasl/sls053}

\end{thebibliography}

\end{document}